\documentclass[11pt]{article}
\usepackage[margin=1in]{geometry}
\usepackage{graphicx, epstopdf}
\usepackage{amsmath, amssymb, amsfonts, bm, mathrsfs, commath}
\usepackage{booktabs, xcolor}
\usepackage{authblk}
\usepackage{multirow}
\usepackage{csquotes}
\usepackage{dsfont}
\usepackage{soul}
\usepackage[T1]{fontenc}
\usepackage{bm}
\usepackage{caption}
\usepackage{booktabs}
\usepackage{doi}
\usepackage{tikz}
\usetikzlibrary{positioning, arrows.meta, shapes.geometric}
\usepackage{algorithm, algpseudocode}
\usepackage{hyperref}
\usepackage{url}
\hypersetup{
	colorlinks   = true, 
	urlcolor     = blue, 
	linkcolor    = blue, 
	citecolor   = blue 
}
\usepackage[authoryear]{natbib}
\usepackage{cleveref}

\title{Climate--Driven Mortality Forecasting Using Deep Learning}

\author[1]{Kenrick So\thanks{Corresponding author: kenrick.so@uclouvain.be}}
\author[1]{Karim Barigou}
\author[2]{Jens Robben}

\affil[1]{Institute of Statistics, Biostatistics and Actuarial Science (ISBA), Louvain Institute of Data Analysis and Modeling (LIDAM), UCLouvain, Louvain-la-Neuve, Belgium.}
\affil[2]{Research Centre for Longevity Risk (RCLR), Faculty of Economics and Business, University of Amsterdam, Amsterdam, The Netherlands.}

\date{}

\begin{document}
\maketitle
\begin{abstract}
    Climate extremes have become important drivers of mortality, producing sudden spikes that traditional mortality models fail to predict. To address this gap, we propose a two-step modelling framework that combines a regional weekly Lee--Carter baseline model that captures long-term mortality trends and overall seasonal patterns, with two complementary deep learning architectures designed to model excess mortality driven by environmental conditions and climate shocks. The first, a CNN--LSTM, captures region-specific temporal responses through convolutional filters. The second, a GNN--LSTM, replaces convolutions with graph-based representations to model spatial mortality dependencies and the propagation of climate-related impacts across regions. Both architectures are further extended to a quantile LSTM framework that produces time-varying prediction intervals. We evaluate our models against both the Lee--Carter baseline and MortFCNet \citep{Zheng2025FineGrained}. Using French regional data over 1990--2019, our models capture delayed and nonlinear associations between environmental extremes and excess mortality. Both proposed architectures outperform the Lee--Carter baseline and MortFCNet across all regions, each reducing test MSE by approximately 24\% relative to the MortFCNet, with particularly large gains at the oldest ages where climate-driven mortality spikes are most severe. From a risk management perspective, the proposed framework provides a more realistic characterization of extreme climate-driven mortality risk, with time-varying prediction intervals that offer a more informed basis for the assessment of climate-related longevity exposure by insurers and pension funds.
\end{abstract}


\section{Introduction}
Life insurers and pension funds rely on mortality models to make financial and regulatory decisions, yet these models are not designed to capture the effects of climate change. Under Solvency II, mortality assumptions directly determine the Solvency Capital Requirement and the valuation of provisions. Pension funds similarly depend on mortality projections to calculate liabilities, assess funding ratios, and meet regulatory requirements. As extreme weather events become more frequent, mortality becomes less predictable, with both immediate and delayed effects. When these dynamics are not captured, both capital requirements and pension liabilities can be misestimated. Despite these implications, climate-related drivers of mortality remain largely absent from standard forecasting models. This raises a key challenge: how can mortality models be adapted to capture these climate-driven effects? We propose two novel deep learning architectures designed to jointly capture temporal dependence, spatial heterogeneity, and climate-driven mortality shocks, and introduce a unified uncertainty quantification framework that integrates parameter, simulation, and residual-based uncertainty to improve mortality forecasting and climate risk management.

\subsection{Climate Risk and Mortality}
Climate risk refers to the negative economic, social, and environmental impacts that arise from climate change. Climate risk can be broadly categorized into transition risk and physical risk \citep{Giglio2021Transition}. Transition risk refers to the potential risks arising from regulatory and policy changes as economies move toward lower greenhouse gas emissions and increased reliance on renewable energy sources \citep{DiFebo2025TransitionRisk}. On the other hand, physical risk represents economic and financial losses resulting from the increasing frequency and severity of extreme weather events, including their long-term impact \citep{Shala2024PhysicalRisk}. Physical risks have been studied in the context of property and casualty insurance, where extreme weather events such as floods lead to losses for insured assets \citep{Calabrese2024PhysicalInsurance}. Beyond property damages, physical risk poses a challenge for life insurance, as extreme weather events increase uncertainty in mortality forecasting. However, the impact of physical risk on mortality remains understudied, as it is difficult to isolate the effect of climate factors from demographic, socioeconomic, and healthcare-related influences \citep{Hainaut2026Socio}. There are established links between climate variables and mortality. The epidemiological literature documents the effects of temperature extremes, humidity, and air quality on death rates, particularly among vulnerable populations such as the elderly \citep{Gasparrini2015dlnmImpact,Robbens2025weeklyEstimation}. Importantly, these effects are not limited to immediate impacts but can unfold over time, motivating the use of models capable of capturing temporal dependencies.
 
\subsection{Limitations of Traditional Mortality Models}
Accurate long-term mortality forecasting is important for the pricing and reserving of longevity-linked liabilities \citep{So2025}. Current mortality forecasting methods assume stable and slowly evolving mortality trends over time. While these methods have demonstrated good performance under regular historical mortality patterns, they are less suited to contexts in which mortality is influenced by non-linear external factors. In particular, climate variability and extreme weather events influence mortality in ways that are seasonal and lagged, causing deviations from long-term trends that are not well captured by past models. Climatic determinants of mortality, despite growing empirical evidence of their importance, remain absent from mortality forecasting models used in actuarial and policy settings. A further limitation of mortality forecasting models is their limited ability to distinguish between structural mortality changes and temporary deviations \citep{vanBerkum2016Structural}. Recent work has attempted to address this limitation by explicitly modelling transient mortality shocks. For instance, \citet{goes2025bayesian} extend the Lee--Carter framework with vanishing jump processes to capture pandemic-related excess mortality while preserving long-term trend dynamics. Such approaches highlight the importance of distinguishing between temporary shocks and structural changes in mortality forecasting. Extreme weather events can generate short-term excess mortality as well as delayed effects, including mortality harvesting and lagged health impacts. Such effects may bias long-term mortality forecasts used in Solvency II and related risk assessment frameworks. Another limitation is that most traditional mortality forecasting models are developed using annual mortality data and therefore cannot exploit the increasingly available weekly mortality records. Recent studies have shown that higher-frequency mortality models can better capture seasonal variation and short-term mortality fluctuations, which are particularly important when assessing the effects of climate and extreme weather events \citep{Begin2025Seasonal,Begin2026Seasonal,Li2026MixedData,Li2026Dynamic,Miao2026GBM}.

\subsection{Deep Learning for Mortality Forecasting}
To move beyond the limitations of traditional mortality models, a growing body of research has applied deep learning to mortality forecasting. An overview of current deep learning approaches to mortality forecasting can be found in \cite{Zheng2026Review}. \citet{Richman2019NNLC} applied multi-population mortality modelling with structure selection, while \citet{Hainaut2018Autoencoder} developed an autoencoder that compresses log-mortality surfaces into latent factors subsequently used for forecasting. More recently, \citet{Wang2023Transformer} proposed a transformer architecture using multi-layer attention and positional encoding, and \citet{Perla2024Multitask} introduced locally connected networks that admit an autoregressive time-series interpretation. Convolutional neural networks (CNNs) have proven effective at capturing structural patterns across age--period mortality surfaces. \citet{Perla2021DeepLC} demonstrated that CNNs can serve as a generalization of the Lee--Carter model, and \citet{Wang2021CNN} introduced a neighborhood-based CNN that flexibly captures nonlinear age, period, and cohort effects. \citet{Zhang2022LSTMCNN} subsequently combined LSTM and CNN layers into an architecture across multiple populations. 

Recurrent neural networks address the temporal limitations of ARIMA-based projection in the Lee--Carter framework, which struggles with short-term shocks and long-range dependencies of the kind generated by extreme weather events. \citet{Chen2023LSTM} introduced long-short term memory (LSTM) neural networks as a replacement for the ARIMA component. More recently, \citet{Shen2024GNN} proposed a graph neural network (GNN) transformer with an adaptive adjacency matrix, capturing relational structure across populations to further improve mortality forecasts. Subsequent work has focused on uncertainty quantification: \citet{Marino2022LSTM} embedded an LSTM within the Lee--Carter framework and generated prediction intervals via Poisson bootstrap, while \citet{Lindholm2022LSTM} trained an LSTM on Lee--Carter residuals under a Poisson specification, improving stability with limited data. Building on these advances, \citet{HyeongChan2022Quantile} has extended LSTM-based mortality models to produce full predictive distributions rather than single-point forecasts with quantile LSTM.

\subsection{Mortality Modelling with Climate Covariates}
Despite this progress, climate effects in mortality forecasting have largely been addressed through traditional statistical models rather than deep learning architectures. The main approach relies on distributed lag non-linear models (DLNMs), which capture delayed environmental exposures \citep{Gasparrini2015dlnmImpact,robben2025penalizeddistributedlagnonlinear}. Recent extensions embed DLNMs within multi-population mortality frameworks to quantify temperature effects \citep{Guibert2025dlnm} or combine them with Lee--Carter structures to jointly model stochastic trends and climate-driven mortality \citep{Min2025dlnmLC}. However, DLNMs struggle when multiple interacting climate variables are considered: multicollinearity and the difficulty of representing cross-variable non-linearities limit their generalizability across regions with heterogeneous climate regimes and demographic structures \citep{Chen2019DLNM}. An alternative strand of literature focuses on extreme climate--mortality dependence. \citet{Li2022} apply extreme value theory to model joint extreme temperature and mortality risk, finding strong extremal dependence between cold-weather events and old-age death counts. In parallel, the actuarial literature has explored more aggregate approaches to incorporating climate information into mortality models. \citet{barigou2025mortality} develop a two-step framework in which seasonal mortality models capture baseline mortality dynamics, and residual deviations are explained using components of the Actuaries Climate Index (ACI) through flexible statistical and machine learning models.

Data-driven frameworks have begun to address these limitations. \citet{Robbens2025weeklyEstimation} propose a two-stage approach in which a regional seasonal baseline is first estimated and then corrected using machine learning models incorporating environmental anomalies, finding temperature-related features informative for short-term mortality deviations. \citet{robben2025granular} develop a regime-switching model with covariate-dependent transition probabilities that distinguishes baseline mortality from heatwave and epidemic shocks. Deep learning offers an extension to DLNMs by jointly learning baseline structure, non-linear environmental interactions, and temporal dependencies within a unified framework. Yet its adoption in climate--mortality modelling remains limited; most neural architectures treat mortality as a function of age and time alone, omitting climate signals. A notable exception is \citet{Zheng2025FineGrained}, who propose a GRU architecture called MortFCNet that directly integrates climate covariates with demographic inputs, demonstrating improved short-term forecasts. This article builds on this direction by extending climate-integrated sequence modelling to a setting with explicit spatial structure, jointly capturing temporal dependence, spatial heterogeneity, and climate-driven shocks within a deep learning framework. We do so through a two-step decomposition that mirrors the emerging literature: a baseline mortality model that captures stable seasonal and age-specific mortality dynamics, while neural networks learn the residual variation driven by climate exposures.

Concretely, this paper makes four contributions. First, we apply the weekly seasonal Lee--Carter model of \citet{robben2025penalizeddistributedlagnonlinear} to capture both age-specific and intra-annual mortality patterns. This overcomes the aggregation limitations of Serfling-type approaches and provides a transparent baseline against which climate-driven mortality deviations can be isolated and compared across age groups and regions. Second, building on this seasonal baseline, we develop a CNN--LSTM architecture that extends the MortFCNet framework of \citet{Zheng2025FineGrained} by incorporating region- and age-specific effects. The CNN extracts cross-regional spatial patterns, while the LSTM captures lagged climate impacts, allowing the model to learn nonlinear relationships between mortality and multiple climate variables. Third, we introduce a novel GNN--LSTM architecture that replaces standard convolutions with graph-based representations of regional adjacency. This enables the model to learn spatial dependencies in climate-related mortality shocks, a particularly important feature given the heterogeneous climate regimes and demographic compositions of French NUTS 2 regions. Fourth, we adapt both deep learning architectures into a quantile LSTM framework by modifying only the output layer to forecast conditional quantiles of the climate-induced residuals. This allows the models to produce time-varying prediction intervals that adjust to changing climate conditions and provides a unified framework for uncertainty quantification.

The remainder of this paper is organized as follows: Section 2 introduces the notation and describes the mortality and climate data, including the construction of weekly death rates, exposure estimates, and environmental covariates from the E-OBS database. Section 3 outlines the modelling framework, beginning with the seasonal Lee--Carter benchmark and then presenting the two neural network models: CNN--LSTM, and GNN--LSTM. Next, Section 4 evaluates the models across the 21 French NUTS 2 regions during the 2015--2019 out-of-sample period, with a focus on forecast accuracy, regional variation, age-specific performance, and learned spatial patterns. Section 5 introduces the quantile LSTM framework used for uncertainty quantification, while Section 6 concludes the article.

\section{Notation and Data} \label{Section: Notation and Data}
We consider a weekly, age-specific mortality model that captures variations across time and regions. Let $d_{a, t, w, r}$ denote the number of deaths occurring in week $w$ of year $t$ for individuals aged $a$ in region $r$. Time is indexed using ISO 8601 weeks, where ISO week 1 begins on the Monday of the week containing the first Thursday of the year. We define $\mathcal{R}$ as the set of regions, $\mathcal{X}$ as the set of age groups, $\mathcal{W}_t$ as the set of weeks, and $\mathcal{T} = \{t_{\min}, t_{\min+1}, \dots, t_{\max}\}$ as the set of years under consideration. To relate deaths to the exposure-to-risk, we define $E_{a,t,w,r}$ as the total number of person-years lived by individuals aged $[a,a+1)$ during week $w$ of year $t$, in region $r$. The observed weekly death rate, interpreted as deaths per unit of exposure, is then given by:
\begin{equation}
    m_{a,t,w,r} = \frac{d_{a,t,w,r}}{E_{a,t,w,r}}.
\end{equation}
We denote by $\mu_{a,t,w,r}$ the weekly force of mortality, representing the instantaneous risk of death and the continuous-time analogue of the observed death rate. Under the common assumption that the force of mortality is approximately constant within each weekly interval, $\mu_{a,t,w,r}$ can be approximated by $m_{a,t,w,r}$.

\subsection{Mortality Data}
Similar to \citet{robben2025penalizeddistributedlagnonlinear}, we utilize individual-level death records from France, obtained from the National Institute of Statistics and Economic Studies (INSEE). Each record includes information on the individual's name, gender, commune of birth, commune of death, date of birth, date of death, and death certificate number. The death records were preprocessed in several steps. Duplicate entries were removed, records with missing birth or death dates were excluded, and observations in which the recorded date of death preceded the date of birth were discarded. Missing or incorrectly specified months or days of death were imputed by sampling from the empirical distribution of deaths over months and days. Following data cleaning, the records were aggregated by five-year age groups, ISO year, ISO week, and administrative region of death. This aggregation required mapping each commune of death to its corresponding department using the first two digits of the postal code, and subsequently mapping departments to administrative regions using the official correspondence table provided by INSEE. For the purposes of this study, we consider the period from 1990 to 2019, focusing on individuals aged 65 and older, grouped into five-year age intervals with an open-ended 90+ category. The analysis is restricted to the 21 NUTS 2 regions of metropolitan France, excluding Corsica and all overseas regions. Corsica is excluded due to its small population size, which leads to sparse death counts, while overseas regions are excluded because their demographic, mortality, and climate patterns differ from those of metropolitan France.

We estimate weekly population exposures by age, sex, and region using annual population counts $P_{a,t,r}$ obtained from INSEE, where $a$ denotes age, $t$ year, and $r$ the administrative region. The counts $P_{a,t,r}$ correspond to the population as of January 1 of year $t$. Weekly population estimates, denoted by $\tilde{P}_{a,t,w,r}$, are constructed via linear interpolation between consecutive annual counts:
\begin{equation}
    \tilde{P}_{a,t,w,r} = P_{a,t,r} + \frac{\mathrm{days}(\mathrm{date}_t, \mathrm{date}_{t,w})}{\mathrm{days}(\mathrm{date}_t, \mathrm{date}_{t+1})} \left( P_{a,t+1,r} - P_{a,t,r} \right),
\end{equation}
where $\mathrm{date}_t$ denotes January 1 of year $t$, $\mathrm{date}_{t,w}$ the start of ISO week $w$ in year $t$, and $\mathrm{days}(\cdot,\cdot)$ returns the number of days between two dates. Weekly exposures $E_{a,t,w,r}$, expressed in person-years, are then computed using the midpoint approximation between successive weekly population estimates:
\begin{equation}
    E_{a,t,w,r} = \frac{\tilde{P}_{a,t,w,r} + \tilde{P}_{a,t,w+1,r}}{2 \times 52.18},
\end{equation}
where $\tilde{P}_{a,t,w+1,r}$ is replaced by $\tilde{P}_{a,t+1,1,r}$ when $w+1$ exceeds the number of ISO weeks in year $t$. The scaling factor $52.18$ corresponds to the average number of weeks per year and ensures that exposures are expressed in annualized units.
 
\subsection{Climate Data}
We incorporate environmental covariates relevant to mortality dynamics using high-resolution meteorological data from the Copernicus Climate Data Store (CDS). Meteorological variables are obtained from the E-OBS land-only gridded observational dataset for Europe \citep{eobs2025}. The data are provided on a regular spatial grid with a resolution of $0.10^\circ$ (approximately 10 km) in both longitude and latitude, and are available at a daily frequency. The selection of environmental covariates is guided by epidemiological evidence on the relationship between weather conditions and mortality. Previous studies have shown that minimum, mean, and maximum temperatures exhibit similar predictive power for daily mortality \citep{Barnett2010Temperature}. Associations between mortality and atmospheric moisture, commonly measured using relative humidity, have also been documented \citep{Braga2002Humidity}. The environmental covariates included in the analysis, together with their definitions and measurement units, are summarized in Table~\ref{tab:env_vars}.

\begin{table}[!htb]
\centering
\caption{Environmental covariates aggregated at the NUTS 2 regional level and used as climate inputs for the neural mortality forecasting models.}
\label{tab:env_vars}
\begin{tabular}{ll}
\toprule
\multicolumn{1}{c}{\textbf{Variable}} &
\multicolumn{1}{c}{\textbf{Definition}} \\
\midrule
$T_{\text{Max}}$ & Daily maximum temperature ($^\circ$C) at 2 metres above the surface. \\
$T_{\text{Avg}}$ & Daily average temperature ($^\circ$C) at 2 metres above the surface. \\
$T_{\text{Min}}$ & Daily minimum temperature ($^\circ$C) at 2 metres above the surface. \\
$\text{Hum}$     & Daily average relative humidity (\%) at 2 metres above the surface. \\
$\text{Rain}$    & Daily total precipitation (mm). \\
$\text{Wind}$    & Daily average wind speed (m/s) at 10 metres above the surface. \\
\bottomrule
\end{tabular}
\end{table}

Daily E-OBS environmental variables are aggregated to the NUTS 2 regional level and then converted to ISO-week bins. For each region-week, the daily series are temporally aggregated to weekly values and aligned with the ISO calendar structure used for the mortality data. The climate covariates from Table~\ref{tab:env_vars} are aggregated to a weekly frequency using unweighted period-appropriate statistics: minimum values for $T_{\min}$, maximum values for $T_{\max}$ and precipitation, and means for $T_{\text{avg}}$, humidity, and wind speed.

\section{A Two-Stage Mortality Forecasting Framework}
We develop a two-step framework for modelling and forecasting weekly mortality rates. In the first step, a baseline model is specified to capture the main demographic, temporal, and seasonal patterns in mortality. In the second step, the remaining variation not explained by the baseline is modelled separately using neural networks, allowing us to account for the effects of environmental covariates and extreme climate-related shocks.

\subsection{Stage 1: Baseline Mortality Model}

\subsubsection{Model Specification}
We begin by specifying a baseline weekly seasonal mortality model based on the Lee--Carter framework, adapted to account for regional and age-specific variation. The model is formulated using a negative binomial distribution to accommodate overdispersion in weekly death counts and is estimated by maximum likelihood (see \citet{robben2025penalizeddistributedlagnonlinear}). Assume that the number of deaths in region $r$ for age group $a$ during ISO week $w$ of year $t$ follows a negative binomial distribution:
\begin{equation}
\label{eq:NB}
    D_{a,t,w,r} \sim \text{NB}\Big(E_{a,t,w,r} \cdot \mu_{a,t,w,r}, \phi_{a,r}\Big),
\end{equation}
where $E_{a,t,w,r}$ denotes the exposure-to-risk and $\mu_{a,t,w,r}$ is the weekly force of mortality. The negative binomial specification accounts for overdispersion in weekly death counts arising from sources such as unobserved heterogeneity, extreme mortality shocks, or model misspecification not absorbed by the mean structure. The dispersion parameter $\phi_{a,r}$ is specified on the logarithmic scale as:
\begin{equation}
    \phi_{a,r} = \exp(\phi_a + \phi_r),
\end{equation}
where $\phi_a$ captures age-specific effects and $\phi_r$ captures region-specific changes in the dispersion. The weekly seasonal baseline mortality model is given by:
\begin{equation} 
    \ln(\mu^{\text{base}}_{a,t,w,r}) = \alpha_{a,r} + \beta_a \kappa_{t,r} + \gamma_a \lambda_{w,r},
    \label{eq:wlc_model}
\end{equation}
where $\alpha_{a,r}$ represents the baseline log-mortality level for age group $a$ in region $r$. The term $\kappa_{t,r}$ is a region-specific annual index capturing long-term mortality dynamics, with $\beta_a$ measuring the common age-specific sensitivity across regions to these trends. The term $\lambda_{w,r}$ captures within-year seasonal variation, with $\gamma_a$ governing age-specific seasonal effects. This specification extends the Lee--Carter framework by separating long-term temporal dynamics, $\beta_a \kappa_{t,r}$, from intra-year seasonal variation, $\gamma_a \lambda_{w,r}$, while maintaining parsimony through shared age-specific loadings \citep{Kleinow2015commonAgeEffect}.

\subsubsection{Forecasting with Structural Changes} \label{Section: Forecasting}
Mortality improvements over time are rarely smooth or linear, and historical data often exhibit sudden shifts due to changes in healthcare, and public health interventions. Consequently, simple random-walk forecasts that assume a constant drift over a long training period effectively draw a straight line between the first and last observed index values, which may not reflect the true evolution of mortality over decades and may be overly sensitive to the choice of calibration window \citep{Enchev2017Structural}. To address this issue, we apply the structural change approach of \citet{vanBerkum2016Structural} to the common factor $K_{t}$,  which enters each region-specific index $\kappa_{t,r}$ as follows:
\begin{equation}
    \kappa_{t,r} = K_t + u_{t,r}, \quad 
    \sum_{r \in \mathcal{R}} u_{t,r} = 0 \quad \forall\, t.
\end{equation}
Here, $K_t$ represents a single national time series capturing the common mortality trend, while $u_{t,r}$ captures region-specific deviations from this trend. The zero-sum constraint implies that these deviations are defined relative to the national average and therefore represent purely cross-regional variation at each point in time. The decomposition is based on the set of regional indices $\kappa_{t,r}$ obtained from the fitted weekly Lee--Carter model, estimated on the full training sample. For each time $t$, the national factor is defined as the cross-regional average
\begin{equation}
    K_t = \frac{1}{|\mathcal{R}|} \sum_{r \in \mathcal{R}} 
    \kappa_{t,r},
\end{equation}
which ensures that the zero-sum constraint is satisfied by construction. The 
regional deviations are then computed as
\begin{equation}
    u_{t,r} = \kappa_{t,r} - K_t,
\end{equation}
so that they correspond to demeaned regional indices and sum to zero across 
regions for each $t$. The national trend may exhibit structural changes over time. Following \citet{vanBerkum2016Structural}, we allow for multiple structural changes in the drift of $K_t$ by applying the methodology of \citet{Bai2003Structural} to the first-order difference of the annual national mean fitted log-mortality series:
\begin{equation}
    \Delta K_t = K_t -  K_{t-1}.
\end{equation}
The optimal number of breakpoints and their locations are selected by minimizing the Bayesian information criterion over candidate piecewise-constant drift specifications, subject to a minimum segment length three to prevent short regimes. Change-point detection is applied only to the national factor, as structural shifts are expected to primarily affect the common mortality trend, while the regional deviations capture shorter-term, mean-reverting local fluctuations. Once the most recent breakpoint $b_{m^*}$ is identified, the implied drift of $K_t$ is estimated by ordinary least squares over the post-break segment $b_{m^*} < t \leq t_{\text{max}}$:
\begin{equation}
    \pi = \frac{K_{t_\text{max}} - K_{b_{m^*+1}}}
    {t_\text{max} - b_{m^*+1}},
\end{equation}
where $m^{*}$ is the index of the last detected break point. This ensures that the forecast reflects the most recently identified structural regime. The national factor is then projected using a random walk with drift:
\begin{equation}
    K_t = K_{t-1} + \pi + \varepsilon_t, \quad 
    \varepsilon_t \sim N(0,\, \sigma^2),
\end{equation}
where the innovations $\varepsilon_t$ capture stochastic deviations from the drift and are assumed to be independent. The change-point derived drift $\pi$ takes precedence over any drift estimated from the full history of $K_t$, ensuring that projections reflect the most recently identified structural regime rather than a long-run average dynamic. Regional deviations $u_{t,r}$ are modelled as independent stochastic processes reflecting region-specific deviations from the national trend. Unlike the national factor, we do not apply change-point detection to these regional deviations; instead, the full historical series is used to estimate their dynamics. For each region $r$, the deviations are assumed to follow an autoregressive process of order one with an intercept \citep{Li2005ACF}:
\begin{equation}
    u_{t,r} = \rho_r + \psi_r u_{t-1,r} + \eta_{t,r}, \quad
    \eta_{t,r} \sim \mathcal{N}(0, \sigma_r^2),
\end{equation}
where $\rho_r$ is a region-specific intercept, $\psi_r$ is the autoregressive coefficient governing the speed of mean-reversion towards the national trend, and $\sigma_r^2$ is the innovation variance quantifying the magnitude of regional shocks. The three parameters are estimated jointly within each region using conditional maximum likelihood, while the estimation procedure is carried out separately for each region. The stationarity condition $|\psi_r| < 1$ is satisfied for all 21 NUTS 2 regions, ensuring that regional mortality patterns do not diverge from the common national trend. Combining the projected national factor and regional deviations yields the forecasted latent mortality indices $h$ years ahead:
\begin{equation}
    \hat{\kappa}_{t_{\text{max}}+h,r} = K_{t_{\text{max}}+h} + \hat{u}_{t_{\text{max}}+h,r}.
\end{equation}
Finally, these forecasted indices are used to predict weekly mortality rates by  combining them with the estimated age-specific coefficients and seasonal effects from the fitted weekly Lee--Carter model:
\begin{equation}
    \log \hat{\mu}^{\text{base}}_{a,t_{\text{max}}+h,w,r} = \hat{\alpha}_{a,r} + \hat{\beta}_a\, 
    \hat{\kappa}_{t_{\text{max}}+h,r} + \hat{\gamma}_a\, \hat{\lambda}_{w,r}
\end{equation}
for each age $a$, week $w$ of year $t_{\text{max}}+h$, and region $r$.

\subsection{Stage 2: Neural Network Models for Climate-Driven Excess Mortality} \label{sec: stage 2}
While the baseline weekly mortality model captures long-term and seasonal trends, it does not account for short-term excess mortality arising from external factors such as disease outbreaks, increased air pollution levels, or extreme weather events. Here we focus on extreme weather as the primary driver, though the framework extends naturally to other data sources. We capture the resulting excess mortality through the log-mortality residuals, defined as the difference between observed and baseline-implied log death rates:
\begin{equation} \label{eq: residuals}
    R_{a,t,w,r} = \log\left(\frac{d_{a,t,w,r}}{E_{a,t,w,r}}\right) - \log (\hat{\mu}^{\text{base}}_{a,t,w,r}),
\end{equation}
where $d_{a,t,w,r}$ and $E_{a,t,w,r}$ denote the observed deaths and exposure-to-risk for age group $a$ in week $w$ of year $t$ in region $r$, and $\hat{\mu}_{a,t,w,r}^{\text{base}}$ is the estimated weekly force of mortality from the baseline model \eqref{eq:wlc_model}. A positive residual $R_{a,t,w,r} > 0$ indicates that observed mortality exceeds what the baseline model predicts, suggesting the presence of excess mortality potentially attributable to climate covariates. Conversely, $R_{a,t,w,r} < 0$ reflects lower-than-expected mortality. Building on this residual-based decomposition, we employ neural network architectures to learn the nonlinear association between the estimated residuals and the region-specific weather-related covariate vectors. The resulting estimates of climate-induced excess mortality, $\hat{R}_{a,t,w,r}$, are then incorporated as an adjustment to the baseline Lee--Carter framework:
\begin{equation}
    \log(\hat{\mu}^{\text{final}}_{a,t,w,r}) = \log(\hat{\mu}^{\text{base}}_{a,t,w,r}) + \hat{R}_{a,t,w,r}.
\end{equation}
In the remainder of this section we index weekly observations by a single time index $\tau$, where $\tau$ corresponds to calendar week $(t,w)$. For notational simplicity, we use $\tau$ throughout and suppress the mapping $(t,w) \leftrightarrow \tau$. For instance, we define $\mathbf{x}_{\tau,r} \in \mathbb{R}^d$ as the vector of $d$ climate features for region $r$ at week $\tau$ (see Table~\ref{tab:env_vars}). We suppress age dependency since covariates are identical across age groups.

\subsubsection{MortFCNet}

As a benchmark we consider MortFCNet \citep{Zheng2025FineGrained}, a residual-based mortality model that estimates weekly mortality residuals from a short window of recent covariates. At each week $\tau$, the model takes as input the covariate sequence $\{\mathbf{x}_{\tau-1,r},\, \mathbf{x}_{\tau,r}\}$. Following their framework, age was not used as an input feature. The inputs are processed by a gated recurrent unit (GRU), which encodes the short covariate window into a latent state. Starting from an initial zero state, the hidden representation is updated according to:
\begin{equation}
\label{eq: GRU}
    \mathbf{h}_{\tau,r} =
    \mathcal{F}\left(\mathbf{h}_{\tau-1,r},\,
    \mathbf{x}_{\tau,r}\right),
\end{equation}
where $\mathcal{F}$ denotes the GRU transition function. In particular, the previous hidden state $\mathbf{h}_{\tau-1,r}$ is obtained by processing the lagged covariate vector $\mathbf{x}_{\tau-1,r}$, so the final state $\mathbf{h}_{\tau,r}$ summarizes information from both weeks. This two-step recurrence produces a latent state $\mathbf{h}_{\tau,r}$ that captures recent covariate dynamics. The latent state $\mathbf{h}_{\tau,r}$ is concatenated with region identifiers $r$, which encodes the region through a categorical integer identifier forming the representation $[\mathbf{h}_{\tau,r}, r]$. This vector is then passed through four fully connected dense layers, where each layer is followed by, respectively, LayerNorm \citep{ba2016layernormalization}, LeakyReLU \citep{Maas2013LeakyReLU}, and dropout \citep{Srivastava2014Dropout} to stabilize gradient flow. We refer to \cite{Zheng2025FineGrained} for further details. This then produces the mortality residual $\hat{R}_{a,\tau,r}$, where $\tau$ refers to a particular year-week combination $(t,w)$. 

\subsubsection{CNN--LSTM}
While MortFCNet captures temporal dependence in mortality residuals through a recurrent structure, it treats the covariates at each time point $\tau$ as a flat input vector and restricts the encoder to a two-week window. Patterns that develop across multiple weeks, such as the lagged mortality consequences of a sustained heatwave or a prolonged cold spell, must therefore be inferred implicitly from the GRU dynamics. To address this limitation, we propose a CNN--LSTM architecture that first extracts explicit local temporal features from the climate sequence, then feeds those features into an LSTM capable of retaining information over much longer horizons.

Let $\mathbf{X}_{\tau, r} = (\mathbf{x}_{\tau - T,r}, \dots, \mathbf{x}_{\tau,r})\in \mathbb{R}^{(T+1)\times d}$ denote the observed climate history for region $r$ over a look-back window of $T+1 = 4$ weeks ending at the current week $\tau$, as used in Algorithm~\ref{Algorithm: CNN--LSTM}. The full sequence $\mathbf{X}_{\tau, r}$ therefore describes how the climate environment evolves over the observation window, week by week, in region $r$.

A one-dimensional convolutional layer scans the sequence with $J$ learned filters. Each filter spans $m=3$ consecutive weeks and learns a local temporal pattern. At position $s$, filter $j$ combines the climate observations within the corresponding $m$-week window into a single feature value:
\begin{equation}
\label{eq: conv}
    z_{s,j}(\mathbf{X}_{\tau, r})
    = g\left(b_j + \sum_{\ell=1}^{m}
      \langle \mathbf{w}_{j,\ell},\, \mathbf{x}_{\tau - T + s+\ell-1,r} \rangle
      \right),
    \qquad s = 0, \dots, S-1,
\end{equation}
where $S = T + 2 - m$ is the number of positions obtained by sliding a width-$m$ filter over the full sequence of length $T+1$, $\mathbf{w}_{j,\ell} \in \mathbb{R}^d$ is a learned weight vector, $b_j$ is a scalar bias, and $g(\cdot)$ is a nonlinear activation function. Note that position $s=0$ covers the earliest $m$ weeks in the window, $\mathbf{x}_{\tau-T,r}, \dots, \mathbf{x}_{\tau-T+m-1,r}$, and position $s=S-1$ covers the most recent $m$ weeks, $\mathbf{x}_{\tau-m+1,r}, \dots, \mathbf{x}_{\tau,r}$, ending at the current week $\tau$. Intuitively, each filter $j$ learns to detect a particular short-term climate pattern, such as a rapid temperature rise or a sustained cold spell, and reports its intensity at each position in the sequence. Stacking the responses of all $J$ filters at position $s$ gives the convolutional feature vector:
\begin{equation*}
    \mathbf{v}_{s,r} = \bigl(z_{s,j}(\mathbf{X}_{\tau,r})\bigr)_{j=1}^{J},
\end{equation*}
computed at line~3 of Algorithm~\ref{Algorithm: CNN--LSTM}. Layer normalization rescales each feature vector to have zero mean and unit variance, stabilizing training, while dropout randomly zeroes out a fraction of activations during training to prevent the model from over-relying on any single feature. After layer normalization and dropout (line~4, Algorithm~\ref{Algorithm: CNN--LSTM}), $\mathbf{v}_{s,r}$ is a $J$-dimensional summary of the climate covariates over the $m$-week window at position $s$. 

Mortality records carry four identifiers: the week, the year, the region $r \in \mathcal{R}$ and the age group $a$. For the region, we define a learnable embedding function \citep{Wuthrich2023Book}:
\begin{equation}
\label{eq: embeddings}
    \xi_{\mathcal{R}} : \mathcal{R} \rightarrow
    \mathbb{R}^{q_{\mathcal{R}}},
\end{equation}
where $q_{\mathcal{R}}$ is the embedding dimension, which maps each region to a learned vector representation of dimension $q_{\mathcal{R}} < |\mathcal{R}|$ that is updated during training. Each region is mapped to a low-dimensional learned vector that is updated during training to capture region-specific mortality characteristics. Rather than assigning a categorical embedding to the age $a$, we represent it as a single standardized scalar:
\begin{equation}
\label{eq: age scalar}
    \tilde{a} = \frac{a_{\mathrm{mid}} - \mu_{a}}{\sigma_{a}},
\end{equation}
where $a_{\mathrm{mid}}$ is the numeric midpoint of the age bin and $\mu_{a}$, $\sigma_{a}$ are the mean and standard deviation of the age midpoints computed over the training set. Given the distributional assumption in Equation~\eqref{eq:NB}, the model parameters are estimated by maximizing the corresponding negative binomial log-likelihood over all observations. The age group is represented as a single standardized scalar, allowing the model to capture the smooth increase of mortality risk with age without requiring a separate embedding per age group. This allows the network to interpolate between observed age groups, and reduces the parameter count relative to a full categorical embedding. The embedding $\xi_{\mathcal{R}}(r)$ and the scalar $\tilde{a}$ are computed once per forward pass at line~1 of Algorithm~\ref{Algorithm: CNN--LSTM}, then appended to the climate feature vector at every convolutional position $s$ to form the LSTM input (line~5, Algorithm~\ref{Algorithm: CNN--LSTM}):
\begin{equation*}
    \mathbf{u}_{s,r,a}
    = [\mathbf{v}_{s,r};\; \xi_{\mathcal{R}}(r);\; \tilde{a}],
\end{equation*}
Appending the region embedding and age scalar at every step means that the LSTM recurrence operates on a sequence in which every position carries explicit information about which subgroup is being modelled, enabling the gating mechanisms to learn region- and age-specific temporal dynamics. 

To capture how the extracted local patterns accumulate and evolve over the observation window, the sequence $(\mathbf{u}_{s,r,a})_{s=0}^{S-1}$ is processed by an LSTM network, corresponding to lines~7--10 of Algorithm~\ref{Algorithm: CNN--LSTM}. Unlike the GRU in MortFCNet, the LSTM maintains both a hidden state $\mathbf{h}_{s,r,a}$ and a memory cell $\mathbf{c}_{s,r,a}$, each an $h$-dimensional vector, where $h$ denotes the number of LSTM units. The hidden state and memory cell are initialized as $\mathbf{h}_{0,r,a}=\mathbf{0}$ and $\mathbf{c}_{0,r,a}=\mathbf{0}$. For $s=0,\ldots,S-1$, the LSTM transition function $\mathcal{L}(\mathbf{u}_{s,r,a}, \mathbf{h}_{s,r,a}, \mathbf{c}_{s,r,a})$ maps the current input, hidden state, and memory cell to the updated states $(\mathbf{h}_{s+1,r,a}, \mathbf{c}_{s+1,r,a})$ through the following gating equations:
\begin{align}
\label{eq: LSTM}
    \mathbf{f}_{s,r,a} &= \text{sig}\left(\mathbf{b}^{(f)}
                   + \mathbf{W}^{(f)} \mathbf{u}_{s,r,a}
                   + \mathbf{V}^{(f)} \mathbf{h}_{s,r,a}\right), \\
    \mathbf{i}_{s,r,a} &= \text{sig}\left(\mathbf{b}^{(i)}
                   + \mathbf{W}^{(i)} \mathbf{u}_{s,r,a}
                   + \mathbf{V}^{(i)} \mathbf{h}_{s,r,a}\right), \\
    \mathbf{o}_{s,r,a} &= \text{sig}\left(\mathbf{b}^{(o)}
                   + \mathbf{W}^{(o)} \mathbf{u}_{s,r,a}
                   + \mathbf{V}^{(o)} \mathbf{h}_{s,r,a}\right), \\
    \tilde{\mathbf{h}}_{s,r,a} &= \tanh\left(\mathbf{b}^{(c)}
                   + \mathbf{W}^{(c)} \mathbf{u}_{s,r,a}
                   + \mathbf{V}^{(c)} \mathbf{h}_{s,r,a}\right), \\
    \mathbf{c}_{s+1,r,a} &= \mathbf{f}_{s,r,a} \odot \mathbf{c}_{s,r,a}
                   + \mathbf{i}_{s,r,a} \odot \tilde{\mathbf{h}}_{s,r,a}, \label{eq: LSTM-Update} \\
    \mathbf{h}_{s+1,r,a} &= \mathbf{o}_{s,r,a} \odot \tanh(\mathbf{c}_{s+1,r,a}),
\end{align}
where $\text{sig}(\cdot)$ is the sigmoid function, $\odot$ denotes element-wise multiplication, and the weight matrices $\mathbf{W}^{(\cdot)} \in \mathbb{R}^{h \times (J + q_{\mathcal{R}} + 1)}$ and $\mathbf{V}^{(\cdot)} \in \mathbb{R}^{h \times h}$ together with bias vectors $\mathbf{b}^{(\cdot)} \in \mathbb{R}^h$ are all learned during training. Each of $\mathbf{f}_s$, $\mathbf{i}_s$, and $\mathbf{o}_s$ is an $h$-dimensional vector with entries in $(0,1)$, corresponding to the forget, input, and output gates. In short, the forget gate decides what to discard from memory, the input gate what to add, and the output gate what to reveal. More specifically, the candidate update $\tilde{\mathbf{h}}_{s,r,a}$ is an $h$-dimensional vector of values in $(-1,1)$, representing a proposed update to the cell state. The cell update $\mathbf{c}_{s+1,r,a}$ in Equation~\eqref{eq: LSTM-Update} is the key distinction from the GRU: the forget gate $\mathbf{f}_{s,r,a}$ selectively erases entries of the memory cell $\mathbf{c}_{s,r,a}$, while the input gate $\mathbf{i}_{s,r,a}$ controls how much of the candidate update is written into memory. This separation of erasing and writing allows the LSTM to maintain long-horizon signals, such as the cumulative mortality impact of a multi-week heat wave, without interference from the short-term noise that tends to destabilize GRU hidden states over longer sequences. The output gate $\mathbf{o}_{s,r,a}$ then finally regulates which memory contents are exposed as the hidden state $\mathbf{h}_{s,r,a}$. 

Let $\mathbf{h}_{S,r,a}$ denote the hidden state produced after processing 
the final convolutional input $\mathbf{u}_{S-1,r,a}$, i.e.\ after the LSTM 
has seen the full observation window up to week $\tau$. This vector encodes the temporal history of the convolutional climate features as an $h$-dimensional vector. After processing the full sequence, $\mathbf{h}_{S,r,a}$ is normalized (line~11, Algorithm~\ref{Algorithm: CNN--LSTM}) and concatenated once more with the region embedding and standardized age scalar (line~12, Algorithm~\ref{Algorithm: CNN--LSTM}):
\begin{equation*}
    \mathbf{z}_{r,a} = [\mathbf{h}_{S,r,a};\; \xi_{\mathcal{R}}(r);\;
    \tilde{a}].
\end{equation*}
This second concatenation, visible in line~12 of Algorithm~\ref{Algorithm: CNN--LSTM}, ensures that the cross-sectional identity of the observation is reinforced at the output stage, complementing the recurrent dynamics that already processed the embedding-augmented sequence $(\mathbf{u}_{s,r,a})_{s=0}^{S-1}$. Similar to MortFCNet, the combined vector $\mathbf{z}_{r,a}$ is then passed through three fully connected dense layers, where each layer is followed by, respectively, LayerNorm, LeakyReLU, and dropout to stabilize gradient flow (lines~13--15, Algorithm~\ref{Algorithm: CNN--LSTM}), and finally through a linear output layer, i.e., $\mathrm{Dense}_1(\cdot)$, to produce the mortality residual at time $\tau$, age group $a$, and region $r$: $\hat{R}_{a,\tau,r} = \mathrm{Dense}_1(\mathbf{z}_{r,a})$ (line~16, Algorithm~\ref{Algorithm: CNN--LSTM}).

\begin{algorithm}[thb!]
    \caption{CNN--LSTM Forward pass at time $\tau$}
    \label{Algorithm: CNN--LSTM}
    \begin{algorithmic}[1]

    \Require \parbox[t]{.85\linewidth}{
Covariate sequence $\mathbf{X}_{\tau,r} = (\mathbf{x}_{\tau - T,r}, \dots, \mathbf{x}_{\tau,r})$ for region $r \in \mathcal{R}$\\
Standardized age scalar $\tilde{a} \in \mathbb{R}$
}

    \Ensure  Mortality residual prediction $\hat{R}_{a,\tau,r}$

    \State Compute $\xi_{\mathcal{R}}(r)$
           \Comment{Learnable embedding, Eq.~\eqref{eq: embeddings}}

    \For{$s = 0$ \textbf{to} $S - 1$} 
        \State $\mathbf{v}_{s,r} \leftarrow \bigl(z_{s,j}(\mathbf{X}_{\tau,r})\bigr)_{j=1}^{J}$
        \Comment{Convolutional features, Eq.~\eqref{eq: conv}}

        \State $\mathbf{v}_{s,r} \leftarrow
        \mathrm{Dropout}\bigl(\mathrm{LeakyReLU}(\mathrm{LayerNorm}(\mathbf{v}_{s,r}))\bigr)$

        \State $\mathbf{u}_{s,r,a} \leftarrow
        [\mathbf{v}_{s,r};\; \xi_{\mathcal{R}}(r);\; \tilde{a}]$
        \Comment{Feature concatenation}
    \EndFor

    \State Initialize $\mathbf{h}_{0,r,a} = \mathbf{0},\; \mathbf{c}_{0,r,a} = \mathbf{0}$

    \For{$s = 0$ \textbf{to} $S - 1$}
        \State $(\mathbf{h}_{s+1,r,a}, \mathbf{c}_{s+1,r,a}) \leftarrow
        \mathcal{L}(\mathbf{u}_{s,r,a}, \mathbf{h}_{s,r,a}, \mathbf{c}_{s,r,a})$
        \Comment{LSTM transition}
    \EndFor

    \State $\mathbf{h}_{S,r,a} \leftarrow
    \mathrm{Dropout}(\mathrm{LayerNorm}(\mathbf{h}_{S,r,a}))$

    \State $\mathbf{z}_{r,a} \leftarrow
    [\mathbf{h}_{S,r,a};\; \xi_{\mathcal{R}}(r);\; \tilde{a}]$

    \For{$l=1$ \textbf{to} 3}
        \Comment{3 dense layers}
        \State $\mathbf{z}_{r,a} \leftarrow
        \mathrm{Dropout}(\mathrm{LeakyReLU}(\mathrm{LayerNorm}(\mathrm{Dense}_{l}(\mathbf{z}_{r,a}))))$
    \EndFor

    \State $\hat{R}_{a,\tau,r} \leftarrow \mathrm{Dense}_1(\mathbf{z}_{r,a})$ \\
    \Return $\hat{R}_{a,\tau,r}$

    \end{algorithmic}
\end{algorithm}

\subsubsection{GNN--LSTM}
While the CNN--LSTM captures local temporal patterns in the climate sequence and incorporates region and age information through a learned embedding and a standardized continuous scalar respectively, it treats each region as an independent unit. Any spatial structure, such as the tendency for mortality shocks to propagate from one region to its neighbors during a widespread heatwave, must be inferred implicitly by the network from the data alone. The GNN--LSTM addresses this directly by replacing the convolutional feature extractor with a graph convolutional network (GCN) that processes all regions jointly at each week, so that every region's representation is explicitly informed by the climate conditions in its spatial neighborhood before the temporal modelling begins. 

We retain the same observation window as in the CNN--LSTM architecture. We denote the sequence of climate observations over a look-back window of $T+1 = 4$ weeks ending at the current week $\tau$ as $\mathbf{X}_{\tau} = (\mathbf{X}_{\tau-T},\ldots,\mathbf{X}_{\tau})$. Here, the matrix $\mathbf{X}_{\tau-T+s}$ collects the climate observations from all regions in week $\tau-T+s$:
\begin{align*}
\mathbf{X}_{\tau-T+s}
=
\begin{bmatrix}
\mathbf{x}_{\tau-T+s,1}^{\top}\\
\vdots\\
\mathbf{x}_{\tau-T+s,R}^{\top}
\end{bmatrix}
\in \mathbb{R}^{R\times d}, 
\end{align*}
for $s=0,\ldots,T$, and where row $r$ contains the $d$-dimensional climate feature vector for region $r$. Unlike the CNN--LSTM, which processes the climate history of a single region, the GNN--LSTM observes the climate conditions of all regions simultaneously at each temporal position $s$.

We represent the $R=21$ French NUTS 2 regions as nodes in a graph,
with spatial dependencies encoded by a fixed adjacency matrix
$\mathbf{A}\in\mathbb{R}^{R\times R}$. The entries $A_{ij}$ equal
$1$ if regions $i$ and $j$ share a border and $0$ otherwise, with $A_{ii}=0$. To ensure numerical stability during neighborhood aggregation, we use the symmetrically normalized adjacency matrix $\tilde{\mathbf A} = \mathbf D^{-1/2} \mathbf A \mathbf D^{-1/2}$, where $\mathbf D$ is the diagonal degree matrix with $D_{ii}=\sum_j A_{ij}$, the number of neighbors of region $i$.

For each temporal position $s=0,\ldots,T$, the GCN processes the regional climate data $\mathbf{X}_{\tau-T+s}$. Node representations are initialized as $\mathbf P_s^{(0)} = \mathbf X_{\tau-T+s}$, and updated through $L$ graph convolutional layers:
\begin{equation}
\label{eq: graph conv}
\mathbf P_s^{(\ell+1)}
=
g\left(
\tilde{\mathbf A}\,
\mathbf P_s^{(\ell)}
\mathbf W^{(\ell)}
+
\mathbf b^{(\ell)}
\right),
\qquad
\ell=0,\ldots,L-1,
\end{equation}
where $\mathbf W^{(\ell)}$ and $\mathbf b^{(\ell)}$ are trainable weight and bias parameters for layer $\ell$, and $g(\cdot)$ is a nonlinear activation function. Note that the matrix product $\tilde{\mathbf A}\, \mathbf P_s^{(\ell)}$ replaces each region's current representation with a weighted sum of the representations of its neighbors, as specified by the normalized adjacency matrix. The subsequent multiplication by $\mathbf W^{(\ell)}$ then applies a learned linear transformation to the aggregated features. Each pass through Equation~\eqref{eq: graph conv} therefore extends the neighborhood each region can see by one further step: after one layer, region $i$ has access to information from its immediate neighbors; after two layers, from its neighbors' neighbors; and so on. After $L$ layers, the output $\mathbf P_s^{(L)}$ contains spatially-informed node representations for all $R$ regions, with feature dimension determined by the last GNN layer. This entire $L$-layer propagation is denoted as $\text{GCN}(\mathbf{X}_{\tau-T+s}, \tilde{\mathbf A})$ at line~3 of Algorithm~\ref{Algorithm: GNN--LSTM}. 

After the final graph convolution layer $\mathbf P_s^{(L)} \in \mathbb{R}^{R\times q_G}$, where $q_G$ denotes the output dimension of the GCN, the representation of the target region $r$ at temporal position $s$ is obtained by selecting the corresponding row (line 4, Algorithm~\ref{Algorithm: GNN--LSTM}):
\begin{align*}
\mathbf v_{s,r}
=
\left(\mathbf P_s^{(L)}\right)_{r,\bullet}. 
\end{align*}
The $q_G$-dimensional vector $\mathbf v_{s,r}$ is a graph-informed feature vector for region $r$ at position $s$, and incorporates information from both the region itself and its spatial neighborhood. The vector $\mathbf v_{s,r}$ is the GNN--LSTM analogue of the convolutional feature vector introduced in the CNN--LSTM architecture. As before, we append the vector $\mathbf v_{s,r}$ with the region embedding $\xi_{\mathcal R}(r)$ and the standardized age scalar $\tilde a$ (line 5, Algorithm~\ref{Algorithm: GNN--LSTM}). The GCN captures shared structure through regional connectivity, while the learnable embeddings allow the model to account for differences between regions that are not explained by the graph alone:
\begin{align*}
\mathbf u_{s,r,a}
=
[\mathbf v_{s,r};\;
\xi_{\mathcal R}(r);\;
\tilde a].
\end{align*}
The sequence $(\mathbf u_{s,r,a})_{s=0}^{T}$ is then processed by the
same LSTM structure as in the CNN--LSTM. Consequently, the hidden states \(\mathbf h_{s,r,a}\) and memory cells \(\mathbf c_{s,r,a}\) evolve exactly as before (lines 7--10, Algorithm~\ref{Algorithm: GNN--LSTM}). The key difference is in what the LSTM is learning to model: where the CNN--LSTM tracked the temporal evolution of local climate pattern scores for a single region, the GNN--LSTM tracks the temporal evolution of the target region's graph-informed climate representation. The LSTM is responsible for learning how these spatially-aggregated target-region representations evolve over the full observation window, capturing delayed, cumulative, and region-specific climate effects that unfold over multiple weeks.

Following the same output stage as in the CNN--LSTM, the final hidden state $\mathbf{h}_{T+1,r,a}$ is normalized, concatenated with the region embedding and age scalar to form $\mathbf{z}_{r,a}$, passed through three fully connected layers, and finally mapped to the mortality residual via $\hat{R}_{a,\tau,r} = \mathrm{Dense}_1(\mathbf{z}_{r,a})$ (lines 11--16, Algorithm~\ref{Algorithm: GNN--LSTM}).

\begin{algorithm}[thb!]
    \caption{GNN--LSTM Forward pass at time $\tau$}
    \label{Algorithm: GNN--LSTM}
    \begin{algorithmic}[1]
    \Require \parbox[t]{.85\linewidth}{
    Regional climate sequence $\mathbf{X}_{\tau} = (\mathbf{X}_{\tau-T},\ldots,\mathbf{X}_{\tau})$ \\
    Normalized adjacency matrix $\tilde{\mathbf{A}}$ \\
    Region index $r \in \mathcal{R}$ \\
    Standardized age scalar $\tilde{a} \in \mathbb{R}$
    }
    \Ensure  Mortality residual prediction $\hat{R}_{a,\tau,r}$

    \State Compute $\xi_{\mathcal{R}}(r)$
           \Comment{Embedding function, Eq.~\eqref{eq: embeddings}}

    \For{$s = 0$ \textbf{to} $T$}
        \State $\mathbf P_s^{(L)} \leftarrow \text{GCN}(\mathbf{X}_{\tau-T+s}, \tilde{\mathbf A})$
               \Comment{GCN propagation, Eq.~\eqref{eq: graph conv}}
        \State $\mathbf v_{s,r} \leftarrow (\mathbf P_s^{(L)})_{r,\bullet}$
               \Comment{Target-region node representation}
        \State $\mathbf u_{s,r,a} \leftarrow [\mathbf v_{s,r};\;\xi_{\mathcal R}(r);\;
        \tilde a]$
               \Comment{Append region embedding and age scalar}
    \EndFor

    \State Initialize $\mathbf{h}_{0,r,a} = \mathbf{0},\; \mathbf{c}_{0,r,a} = \mathbf{0}$

    \For{$s = 0$ \textbf{to} $T$}
        \State $(\mathbf{h}_{s+1,r,a}, \mathbf{c}_{s+1,r,a}) \leftarrow
        \mathcal{L}(\mathbf{u}_{s,r,a}, \mathbf{h}_{s,r,a}, \mathbf{c}_{s,r,a})$
               \Comment{LSTM update, Eqs.~\eqref{eq: LSTM}}
    \EndFor

    \State $\mathbf{h}_{T+1,r,a} \leftarrow
           \mathrm{Dropout}(\mathrm{LayerNorm}(\mathbf{h}_{T+1,r,a}))$

    \State $\mathbf{z}_{r,a} \leftarrow
           [\mathbf{h}_{T+1,r,a};\; \xi_{\mathcal{R}}(r);\; \tilde{a}]$

    \For{$l=1$ \textbf{to} 3}
    \Comment{3 dense layers}
    \State $\mathbf{z}_{r,a} \leftarrow
           \mathrm{Dropout}\bigl(\mathrm{LeakyReLU}\bigl(
           \mathrm{LayerNorm}(\mathrm{Dense}_{l}(\mathbf{z}_{r,a}))\bigr)\bigr)$
    \EndFor
    
    \State $\hat{R}_{a,\tau,r} \leftarrow \mathrm{Dense}_1(\mathbf{z}_{r,a})$ \\
    \Return $\hat{R}_{a,\tau,r}$
    \end{algorithmic}
\end{algorithm}

\section{Empirical Application: French NUTS 2 Regions}
We apply the proposed framework to the weekly mortality data across 21 metropolitan French NUTS 2 regions and six five-year age groups spanning ages 65 to 90+, as discussed in Section~\ref{Section: Notation and Data}. We begin by describing the two-step model calibration procedure, covering both the estimation of the weekly seasonal Lee--Carter baseline and the training of the neural network models. Next, we detail the cross-validation strategy adopted for hyperparameter selection and model comparison. We then report the predictive performance of the proposed approach using out-of-sample forecasts for the 2015--2019 period, both at the aggregate level and across age groups and regions. Lastly, we examine the spatial patterns captured by the learned region embeddings and investigate the role of climatic drivers through permutation feature importance and temperature--mortality response analyses.

\subsection{Baseline Estimation}
Prior to model fitting, we inspected the INSEE weekly death data for reporting anomalies. Notably, June and July 1997 exhibit an unusually large and implausible drop in reported deaths, attributable to recording errors.\footnote{https://www.data.gouv.fr/datasets/fichier-des-personnes-decedees/discussions} To prevent these anomalies from biasing the estimation of mortality trends and seasonal patterns, we treat the affected weeks as missing during calibration, and estimate the negative-binomial Lee--Carter baseline, outlined in Eqs.~\eqref{eq:NB} and \eqref{eq:wlc_model}, using maximum likelihood on the remaining valid data. The missing weeks are subsequently replaced with the model-predicted baseline death counts, ensuring that the models would not learn spurious mortality spikes.

To ensure identifiability, we impose several constraints. The age-specific log-dispersion terms are constrained to sum to zero:
\begin{equation}
    \sum_{a \in \mathcal{A}} \phi_a = 0,
\end{equation}
while the first age group's scale is fixed by setting $\beta_{a_{\min}} = \gamma_{a_{\min}} = 1$ for $a_{\min} = 65$. Additionally, we set the first year's regional temporal factors to zero, $\kappa_{t_{\min},r} = 0$ for all $r$, as well as the first week's seasonal factors, $\lambda_{w_{\min},r} = 0$, with $t_{\min} = 1990$ and $w_{\min} = 1$. These constraints remove location and scale ambiguities, allowing the temporal and seasonal indices to be interpreted meaningfully. Model parameters are then estimated by maximizing the negative binomial log-likelihood using numerical optimization, and standard errors are obtained from the inverse of the Hessian matrix, providing measures of precision for the age-, region-, temporal-, and seasonal-specific effects. 

\begin{figure}[!htb]
    \centering
    \includegraphics[width=0.70\linewidth]{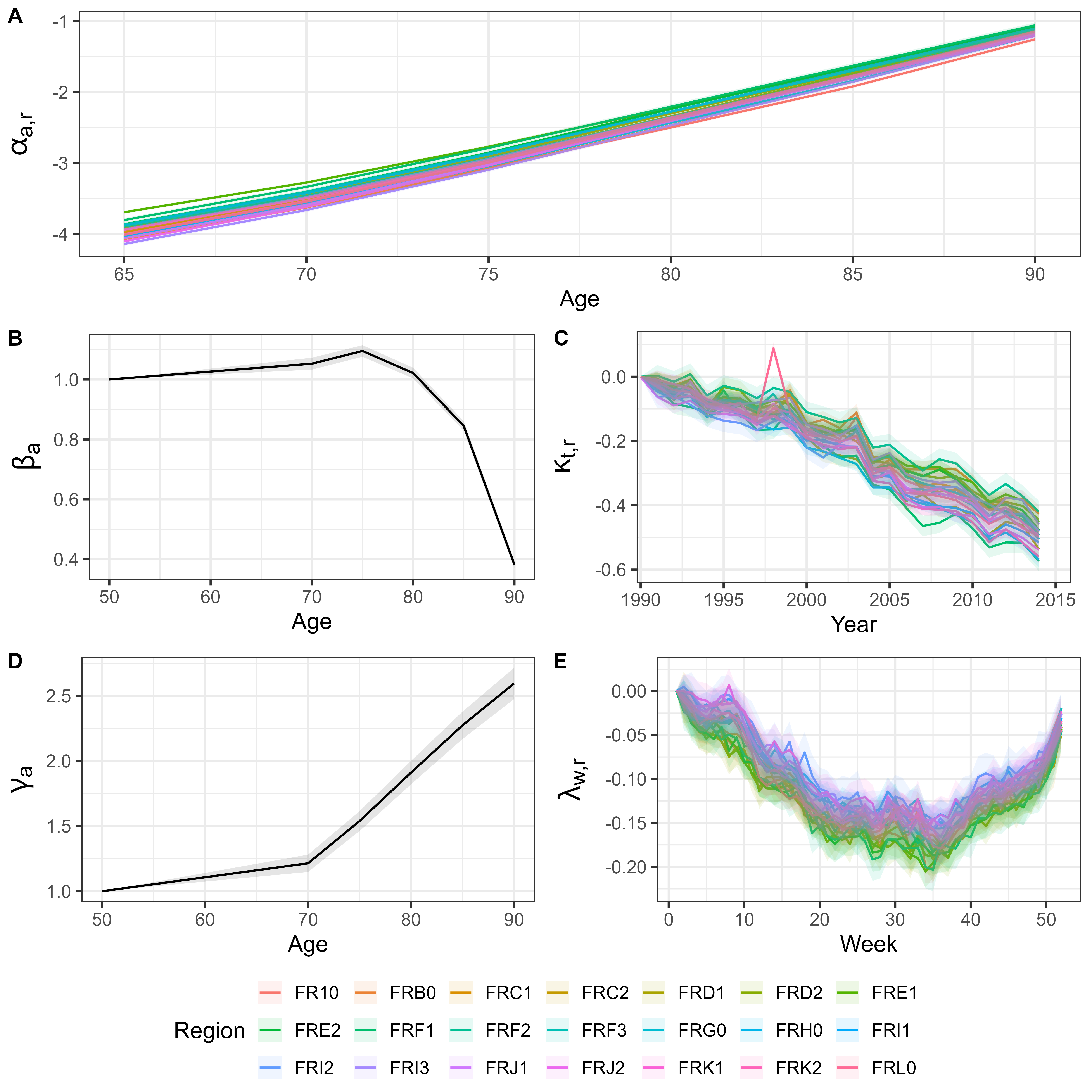}
    \caption{Estimated parameters of the weekly seasonal Lee-Carter model in metropolitan French NUTS 2 regions (1990--2014): $\hat{\alpha}_{a,r}$ (A), $\hat{\beta}_a$ (B), $\hat{\kappa}_{t,r}$ (C), $\hat{\gamma}_a$ (D), $\hat{\lambda}_{w,r}$ (E). 95\%  Confidence intervals are computed from the inverse Hessian of the log-likelihood. Age-specific parameters that are common across regions ($\hat{\beta}_a,\hat{\gamma}_a$) are shown in black, while region-specific parameters ($\hat{\alpha}_{a,r},\hat{\kappa}_{t,r},\hat{\lambda}_{w,r}$) are shown separately by region.}
    \label{fig:LeeCarterPlot}
\end{figure}

Figure~\ref{fig:LeeCarterPlot} visualizes the estimated model parameters for the weekly seasonal Lee-Carter model. Panel A shows the age-level parameters $\hat{\alpha}_{a,r}$, which increase monotonically with age and are tightly grouped across regions, confirming that the overall age pattern of log-mortality is geographically homogeneous. Panel B displays the age-sensitivity loadings $\hat{\beta}_{a}$, which peak around ages 75--80 before declining sharply, indicating that middle-old ages are most responsive to annual mortality improvements. Panel C shows the period index $\hat{\kappa}_{t,r}$, which trends steadily downward from 1990 to 2015 across all regions. Panel D presents $\hat{\gamma}_a$, the age-varying seasonal amplitude, which rises steeply with age, confirming that older individuals face disproportionately larger intra-year mortality fluctuations. Finally, Panel E shows the weekly seasonal pattern $\hat{\lambda}_{w,r}$. We find that all regions exhibit a pronounced U-shaped trough, centered around weeks 25--35.

\subsection{Residual Model Estimation}
After estimating the baseline model, we compute the log-mortality residuals $R_{a,t,w,r}$ in Eq.~\eqref{eq: residuals} as the deviation between observed log-mortality and the baseline prediction for age group $a$, year $t$, week $w$, and region $r$. To explain these residual variations, we use spatio-temporal climate information, where each $\mathbf{x}_{t,w,r}$ is a vector of regional climate covariates observed at week $w$ of year $t$ in region $r$. In addition, the model incorporates information about the region $r$ and age group $a$. A neural network $f_{\boldsymbol{\theta}}(\cdot)$ with parameter vector $\boldsymbol{\theta}$ is trained to predict the residual $R_{a,t,w,r}$ from these inputs, and outputs a predicted residual $\hat{R}_{a,t,w,r}$. Let $\mathcal{D}_{\mathrm{train}}$ denote the set of all tuples $(a,t,w,r)$ in the training period 1990--2014. The model is fitted by minimizing the average squared difference between the observed and predicted residuals:
\begin{equation}
    \mathcal{L}(\boldsymbol{\theta})
    =
    \frac{1}{|\mathcal{D}_{\mathrm{train}}|}
    \sum_{(a,t,w,r)\in\mathcal{D}_{\mathrm{train}}}
    \bigl(R_{a,t,w,r} 
    - f_{\boldsymbol{\theta}}\bigl(\mathbf{x}_{t,r}, r, a\bigr)\bigr)^2.
\end{equation}
To assess predictive performance, we compute the mean squared error on a held-out test set $\mathcal{D}_{\mathrm{test}}$ covering the period 2015--2019. The evaluation metric is:
\begin{equation}
    \mathrm{MSE}
    =
    \frac{1}{|\mathcal{D}_{\mathrm{test}}|}
    \sum_{(a,t,w,r)\in\mathcal{D}_{\mathrm{test}}}
    \bigl(R_{a,t,w,r} - \hat{R}_{a,t,w,r}\bigr)^2,
\end{equation}
where $R_{a,t,w,r}$ and $\hat{R}_{a,t,w,r}$ denote the observed and predicted residuals in the test period, respectively.

\subsection{Cross--Validation} \label{Section: Cross-Validation}
We adopt a cross-validation framework to select hyperparameters for both the Lee--Carter baseline and the neural network models. We therefore partition the data into a training period $\mathcal{T}_{\text{train}} = \{1990,1991,\dots,2011\}$, a validation period $\mathcal{T}_{\text{val}} = \{2012,2013,2014\}$, and a test period $\mathcal{T}_{\text{test}} = \{2015,2016,\dots,2019\}$. Hyperparameter tuning and model selection are performed on $\mathcal{T}_{\text{val}}$, and final out-of-sample evaluation is reserved for $\mathcal{T}_{\text{test}}$, which remains unseen throughout model development. Input sequences span four weeks $\{\mathbf{x}_{t-3,r},\mathbf{x}_{t-2,r},\mathbf{x}_{t-1,r},\, \mathbf{x}_{t,r}\}$ for the CNN--LSTM and GNN--LSTM, and two weeks $\{\mathbf{x}_{t-1,r},\, \mathbf{x}_{t,r}\}$ for MortFCNet, consistent with the configuration in \citet{Zheng2025FineGrained}.

Hyperparameters are selected via grid search across the three neural network architectures. The search space includes learning rates $\{1\times10^{-5}, 5\times10^{-5}, 1\times10^{-4}\}$, LSTM hidden dimensions $\{8,16,32\}$, feedforward network depths $\{32\text{--}16\text{--}8,16\text{--}8\text{--}4,8\text{--}4\text{--}2\}$, dropout rates $\{0.10,0.20\}$, and training epochs $\{30,50,70,150,200\}$. All models are trained with a batch size of 16. The CNN--LSTM uses 16 convolutional filters with a region embedding dimension of 6, while the GNN--LSTM additionally incorporates graph-structured region adjacency with the same embedding scheme and considers GNN hidden dimensions $\{16,32\}$. The MortFCNet is a fully connected residual architecture without region or age embeddings. This results in 1080 hyperparameter configurations in total: 270 for the CNN--LSTM, 540 for the GNN--LSTM, and 270 for the MortFCNet.

Final model selection is based on the mean squared error on the validation set. The selected configurations are as follows. The GNN-LSTM uses GNN and LSTM hidden dimensions of 16, feedforward structure $\{32\text{--}16\text{--}8\}$, a dropout rate of 0.10, and a learning rate of $1\times10^{-4}$. In our implementation, $L=1$ so the GCN consists of a single graph convolutional layer with 16-dimensional output representations. We choose $L=1$ because the graph has a simple structure and we expect neighboring regions to capture the most relevant spatial dependencies. A single graph convolution enables each region to aggregate information from adjacent regions that share a border, while avoiding over smoothing that can arise with deeper GCNs.  The CNN--LSTM uses an LSTM hidden dimension of 16, feedforward structure $\{32\text{--}16\text{--}8\}$, a dropout rate of 0.10, and a learning rate of $5\times10^{-5}$. Lastly, MortFCNet uses an LSTM hidden dimension of 32, feedforward structure $\{8\text{--}4\text{--}2\}$, a dropout rate of 0.20, and a learning rate of $5\times10^{-5}$. Based on cross-validation performance, the number of training epochs is set to 70 for all three neural network architectures, see Table~\ref{tab:epoch_comparison}.

\begin{table}[!htb]
\centering
\small
\renewcommand{\arraystretch}{1.2}
\setlength{\tabcolsep}{4pt}
\begin{tabular}{lcccccc}
\toprule
& \multicolumn{3}{c}{\textbf{Training}} & \multicolumn{3}{c}{\textbf{Validation}} \\
\cmidrule(lr){2-4} \cmidrule(lr){5-7}
\textbf{Epochs} & \textbf{MortFCNet} & \textbf{CNN--LSTM} & \textbf{GNN--LSTM} & \textbf{MortFCNet} & \textbf{CNN--LSTM} & \textbf{GNN--LSTM} \\
\midrule
70 Epochs
& $3.12\times10^{-4}$
& $2.44\times10^{-4}$
& $2.30\times10^{-4}$
& $1.98\times10^{-4}$
& $\boldsymbol{1.61\times10^{-4}}$
& $\boldsymbol{1.56\times10^{-4}}$ \\

200 Epochs
& $\boldsymbol{3.09\times10^{-4}}$
& $\boldsymbol{2.43\times10^{-4}}$
& $\boldsymbol{2.17\times10^{-4}}$
& $1.98\times10^{-4}$
& $1.68\times10^{-4}$
& $1.57\times10^{-4}$ \\
\bottomrule
\end{tabular}
\caption{Training and validation MSE for MortFCNet, CNN--LSTM, and GNN--LSTM models evaluated after 70 and 200 training epochs with the configuration outlined in Section~\ref{Section: Cross-Validation}.}
\label{tab:epoch_comparison}
\end{table}

\subsection{Out-of-Sample Forecast Evaluation}
We evaluate our model performance over the five-year out-of-sample test period $\mathcal{T}_{\text{test}}$. We first report performance through overall training and test losses. To better understand heterogeneity across the population, we then compare results by age group and across regions. Finally, we explore the geographic structure learned by each model through the estimated regional embeddings.

\subsubsection{Aggregate Forecast Performance}
Table~\ref{tab:train_test_mse} shows the overall training and test mean squared errors for all four models. All three models clearly perform better than the Lee--Carter baseline on both the training and test data, confirming that incorporating climate information through machine learning improves predictive accuracy over the traditional approach. 
Among the three neural network architectures, GNN--LSTM achieves the best overall performance, reducing the baseline test error by approximately 24\%. CNN--LSTM achieves a comparable reduction of approximately 24\%, while MortFCNet offers only a 2\% improvement. GNN--LSTM pulls ahead of CNN--LSTM suggesting that explicitly encoding spatial dependencies across regions provides a slight advantage at the aggregate level.

\begin{table}[!htb]
\centering
\renewcommand{\arraystretch}{1.2}
    \begin{tabular}{lcc}
    \toprule
    \textbf{Model} & \textbf{Train MSE} & \textbf{Test MSE} \\
    \midrule
    Baseline  & $3.13\times10^{-4}$ & $1.65\times10^{-4}$ \\
    MortFCNet & $3.03\times10^{-4}$ & $1.62\times10^{-4}$ \\
    CNN--LSTM  & $2.38\times10^{-4}$ & $1.26\times10^{-4}$ \\
    GNN--LSTM  & $\boldsymbol{2.20\times10^{-4}}$ & $\boldsymbol{1.25\times10^{-4}}$ \\
    \bottomrule
    \end{tabular}
\caption{Train and test MSE for the baseline and three neural networks models evaluated over the training period 1990--2014 and test period 2015--2019. The lowest error in each column is highlighted in bold.}
\label{tab:train_test_mse}
\end{table}

\subsubsection{Forecast Performance by Age Group}
Table~\ref{tab:age_group_performance} presents the MSE values computed over the test set in each model. The relative advantage of the neural networks over the Lee--Carter baseline grows steadily with age. At younger ages (65--74), the baseline remains competitive, though both the GNN--LSTM and CNN--LSTM already achieve lower errors. MortFCNet, by contrast, performs worse than the baseline in this range because it assumes that climate covariates have the same effects for all age groups. From age 80 onward, all neural network models begin to outperform the baseline, with the difference growing at each successive age group. Throughout the 75--89 range, the GNN--LSTM and CNN--LSTM track each other closely, with the GNN--LSTM holding a consistent advantage across all three age groups. At 90+, the CNN--LSTM achieves an error of $5.2910 \times 10^{-4}$ against the GNN--LSTM's $5.3024 \times 10^{-4}$, a marginal difference that makes the two architectures essentially equivalent at the oldest age group.

\begin{table}[!htb]
\centering
\renewcommand{\arraystretch}{1.2}
\begin{tabular}{lcccc}
\toprule
\textbf{Age Group} & \textbf{Baseline} & \textbf{MortFCNet} & \textbf{CNN--LSTM} & \textbf{GNN--LSTM} \\
\midrule
65--69 & 6.4048$\times10^{-6}$ & 9.4056$\times10^{-6}$ & 6.1182$\times10^{-6}$ & $\boldsymbol{5.7107\times10^{-6}}$ \\
70--74 & 1.0542$\times10^{-5}$ & 1.3185$\times10^{-5}$ & 1.0131$\times10^{-5}$ & $\boldsymbol{9.8516\times10^{-6}}$ \\
75--79 & 2.3376$\times10^{-5}$ & 2.5275$\times10^{-5}$ & 2.2183$\times10^{-5}$ & $\boldsymbol{2.1593\times10^{-5}}$ \\
80--84 & 5.2279$\times10^{-5}$ & 5.2009$\times10^{-5}$ & 4.7234$\times10^{-5}$ & $\boldsymbol{4.5353\times10^{-5}}$ \\
85--89 & 1.7367$\times10^{-4}$ & 1.6971$\times10^{-4}$ & 1.4399$\times10^{-4}$ & $\boldsymbol{1.3882\times10^{-4}}$ \\
90+    & 7.2158$\times10^{-4}$ & 7.0264$\times10^{-4}$ & $\boldsymbol{5.2910\times10^{-4}}$ & 5.3024$\times10^{-4}$ \\
\bottomrule
\end{tabular}
\caption{Test MSE by age group for the baseline and three neural network models, evaluated over the training period 1990--2014 and test period 2015--2019. The lowest error in each row is highlighted in bold.}
\label{tab:age_group_performance}
\end{table}
\subsubsection{Forecast Performance by Region}
\begin{figure}[!htb]
    \centering
    \includegraphics[width=1\linewidth]{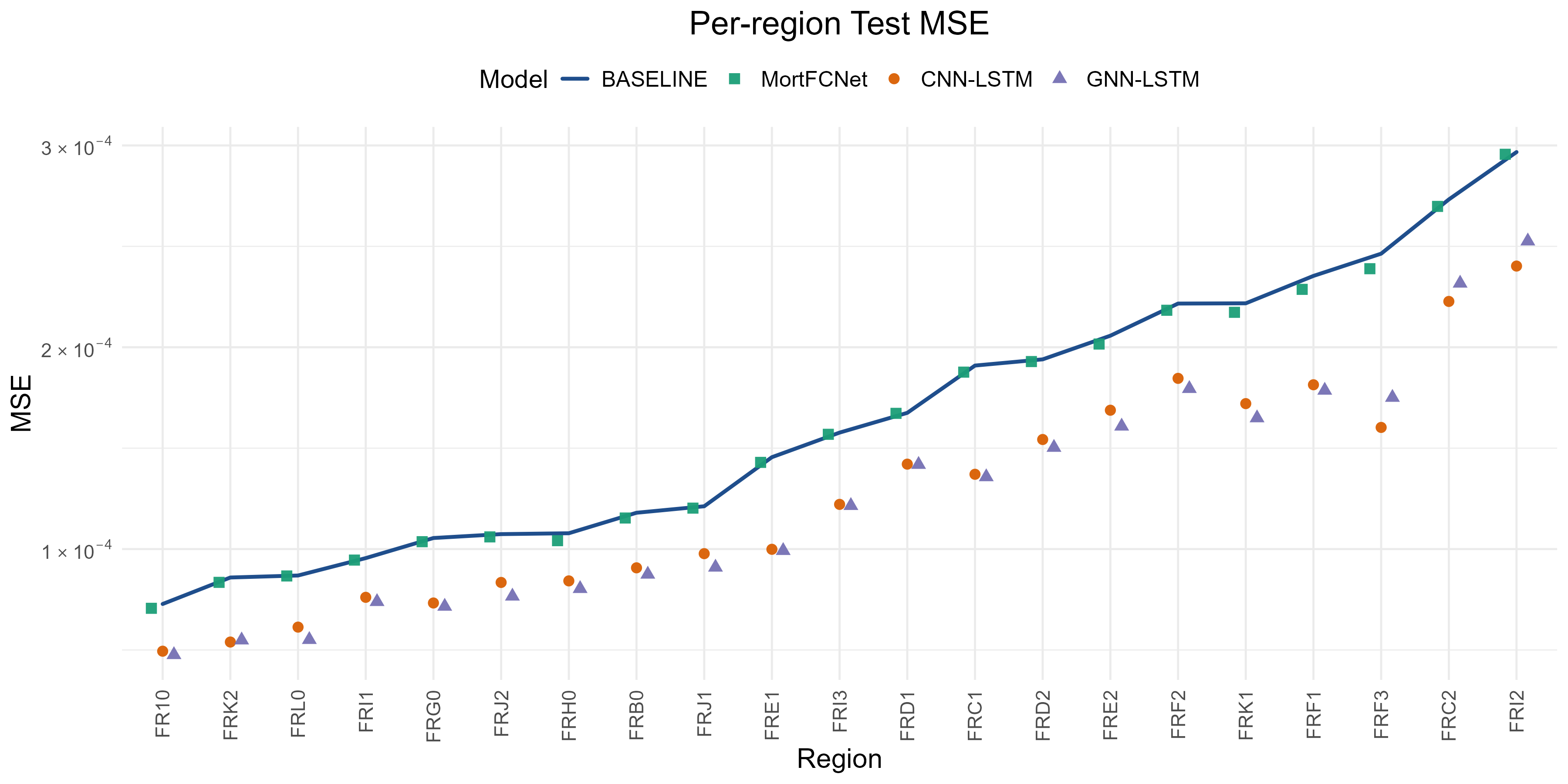}
    \caption{Test MSE by NUTS 2 region, sorted in ascending order of Lee--Carter baseline error. The solid line represents the baseline, while colored points denote the three deep learning models. Points below the baseline line indicate improved predictive performance.}
    \label{fig:per_region_mse}
\end{figure}

To provide a regional view of the model performance, Figure~\ref{fig:per_region_mse} reports the test MSE for each NUTS 2 region. The Lee--Carter model is shown as a connected reference line, while the three deep learning models are displayed as points. Observations below the baseline indicate an improvement in predictive accuracy. MortFCNet closely tracks the baseline across all regions, offering little improvement over the baseline. By contrast, the CNN--LSTM and GNN--LSTM both beat the baseline by a substantial margin across all regions, consistently far below the reference line. A clear pattern emerges between these two architectures: in regions with smaller baseline errors, the GNN--LSTM outperforms the CNN--LSTM, whereas in the regions with the three largest baseline errors, the CNN--LSTM pulls ahead of the GNN--LSTM. This crossover suggests that spatial propagation across neighboring regions is most beneficial where mortality dynamics are relatively smooth and well-structured, while temporal convolutional filters gain the upper hand in regions where mortality is more volatile and harder to predict. 

\begin{figure}[!htb]
    \centering
    \includegraphics[width=1\linewidth]{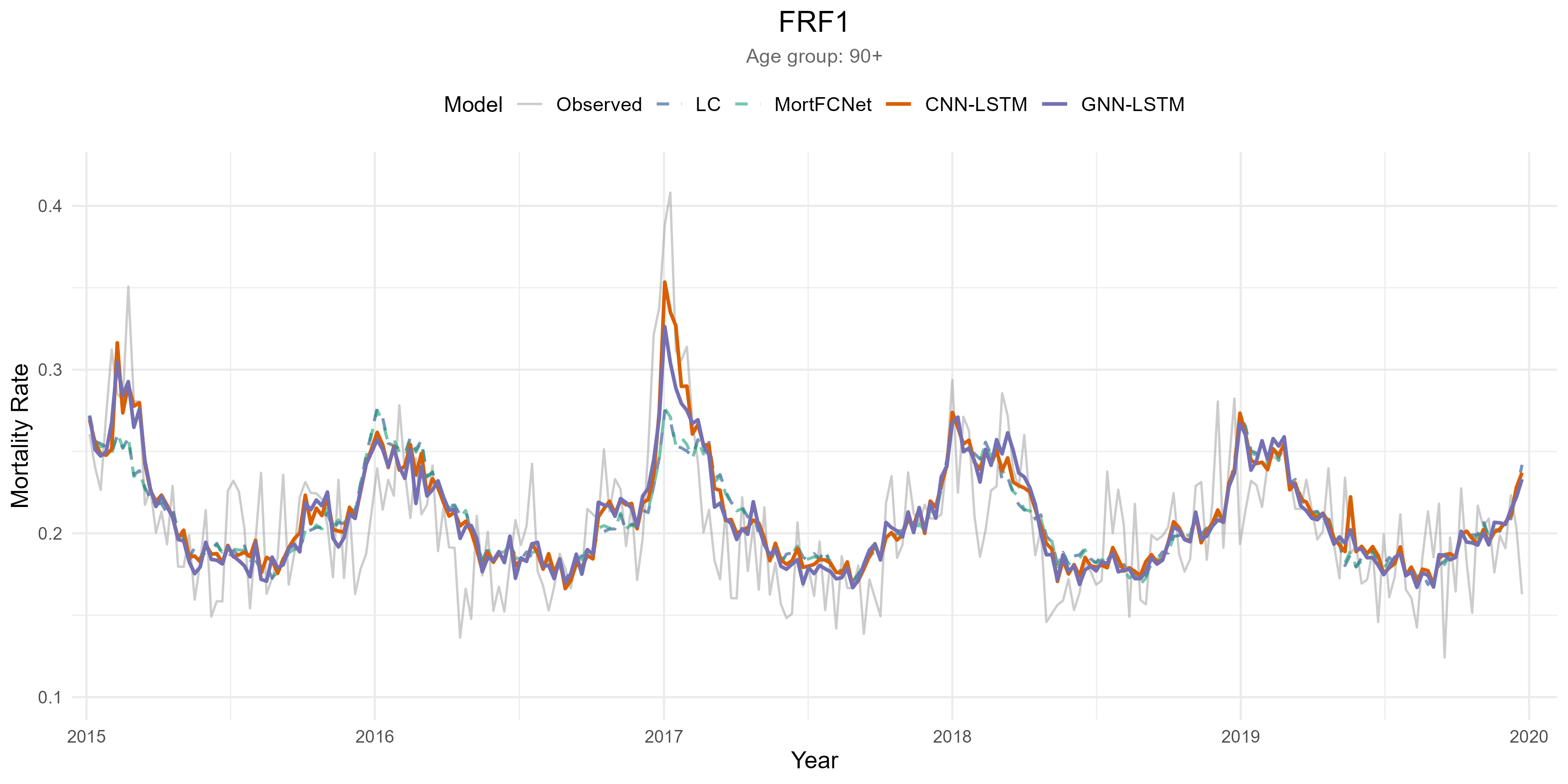}
    \caption{Weekly mortality rate forecasts in Alsace (FRF1), age group 90+, with training period 1990--2014 and test period 2015--2019. Each line corresponds to one of the four models: the weekly seasonal Lee--Carter baseline (blue, dashed), MortFCNet (green, dashed), CNN--LSTM (orange, solid), and GNN--LSTM (purple, solid). The gray line shows the observed mortality rate.}
    \label{fig:top_cnn_frg0}
\end{figure}

\begin{figure}[!htb]
    \centering
    \includegraphics[width=1\linewidth]{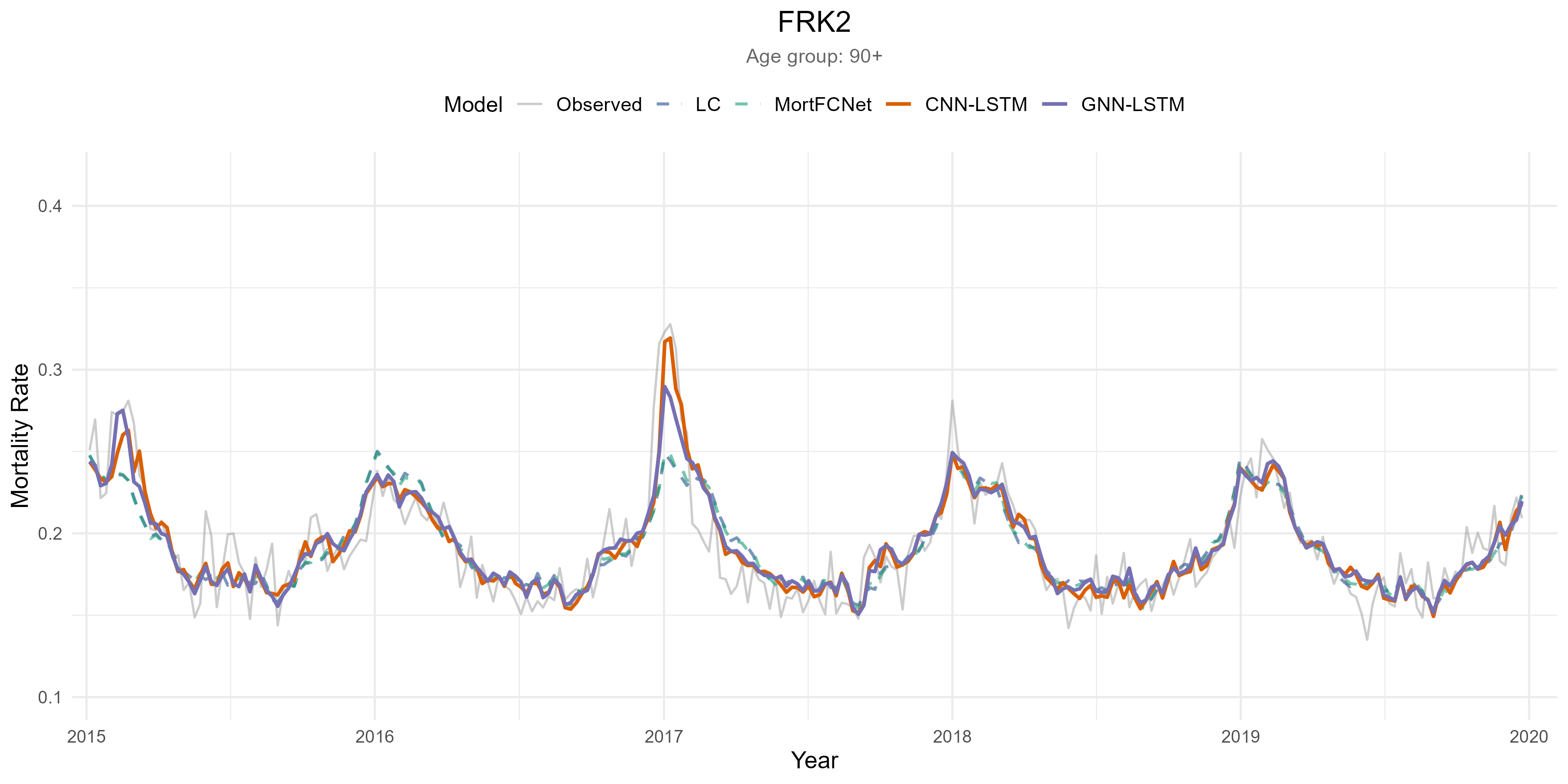}
    \caption{Weekly mortality rate forecasts for region FRK2, age group 90+, with training period 1990--2014 and testing period 2015--2019 . Each line corresponds to one of the four models: the weekly seasonal Lee--Carter baseline, MortFCNet, CNN--LSTM, and GNN--LSTM. The gray line shows the observed mortality rate.}
    \label{fig:top_gnn_frk2}
\end{figure}

To visualize regional forecast performance, Figures~\ref{fig:top_cnn_frg0} and~\ref{fig:top_gnn_frk2} show weekly mortality rate forecasts in Alsace (FRF1) and Rhône-Alpes (FRK2) for the 90+ age group over the 2015--2019 test period. In both regions, the observed series exhibits a seasonal structure: elevated mortality in the first quarter of each year corresponds to winter peaks, while smaller but consistent secondary increases occur during summer months. The Lee--Carter baseline follows the broad seasonal cycle but produces an overly smooth curve that systematically underestimates peak mortality and overestimates trough levels. MortFCNet behaves similarly, with predictions that largely overlap the baseline and show little additional responsiveness to short-term fluctuations in the observed series. By contrast, the CNN--LSTM and GNN--LSTM capture the major mortality peaks throughout the test period. Most notably, both models reproduce the pronounced winter spike in early 2017, during which observed mortality reaches approximately 0.42 in Alsace and 0.33 in Rhône-Alpes, representing the largest mortality peak in the test window. The baseline model and MortFCNet, in contrast, remain largely flat and do not capture this event. Both our of proposed models capture the winter peaks, as well as the moderate summer increases in mortality, where the baseline and MortFCNet do not.

\subsubsection{Regional Embedding Clusters} \label{sec: Regional Embedding Clusters}
The model incorporates a learnable embedding layer, introduced in Section~\ref{sec: stage 2}, that maps each region $r = 1, \dots, R$ to a vector $\mathbf{e}_r := \xi_{\mathcal{R}}(r) \in \mathbb{R}^{q_{\mathcal{R}}}$. These embeddings are learned during training and encode similarities between regions. To visualize this structure, we project the embeddings into two dimensions via Principal Component Analysis (PCA). Let $\mathbf{E} \in \mathbb{R}^{R \times q_{\mathcal{R}}}$ be the matrix of stacked embeddings and $\mathbf{V}_2 \in \mathbb{R}^{q_{\mathcal{R}} \times 2}$ the matrix whose columns are the first two principal component directions of $\mathbf{E}$. The two-dimensional projection of each embedding onto the subspace spanned by the first two principal component directions is then:
\begin{equation}
    \tilde{\mathbf{e}}_r = \mathbf{V}_2^\top \mathbf{e}_r \in \mathbb{R}^2.
\end{equation}
To identify groups of similar regions, we apply $k$-means clustering to the projected embeddings, partitioning the regions into $k$ clusters $\mathcal{C}_1, \dots, \mathcal{C}_k$ by minimizing within-cluster variance:
\begin{equation}
    \min_{\mathcal{C}_1, \dots, \mathcal{C}_k}
    \sum_{i=1}^k \sum_{r \in \mathcal{C}_i}
    \left\| \tilde{\mathbf{e}}_r - \boldsymbol{\mu}_i \right\|^2,
\end{equation}
where $\boldsymbol{\mu}_i$ denotes the centroid of cluster $i$. We select $k=4$, and overlay the resulting cluster assignments on the PCA-reduced coordinates for visualization for both CNN-LSTM and GNN-LSTM.

\begin{figure}[!htb]
    \centering
    \includegraphics[width=1\linewidth]{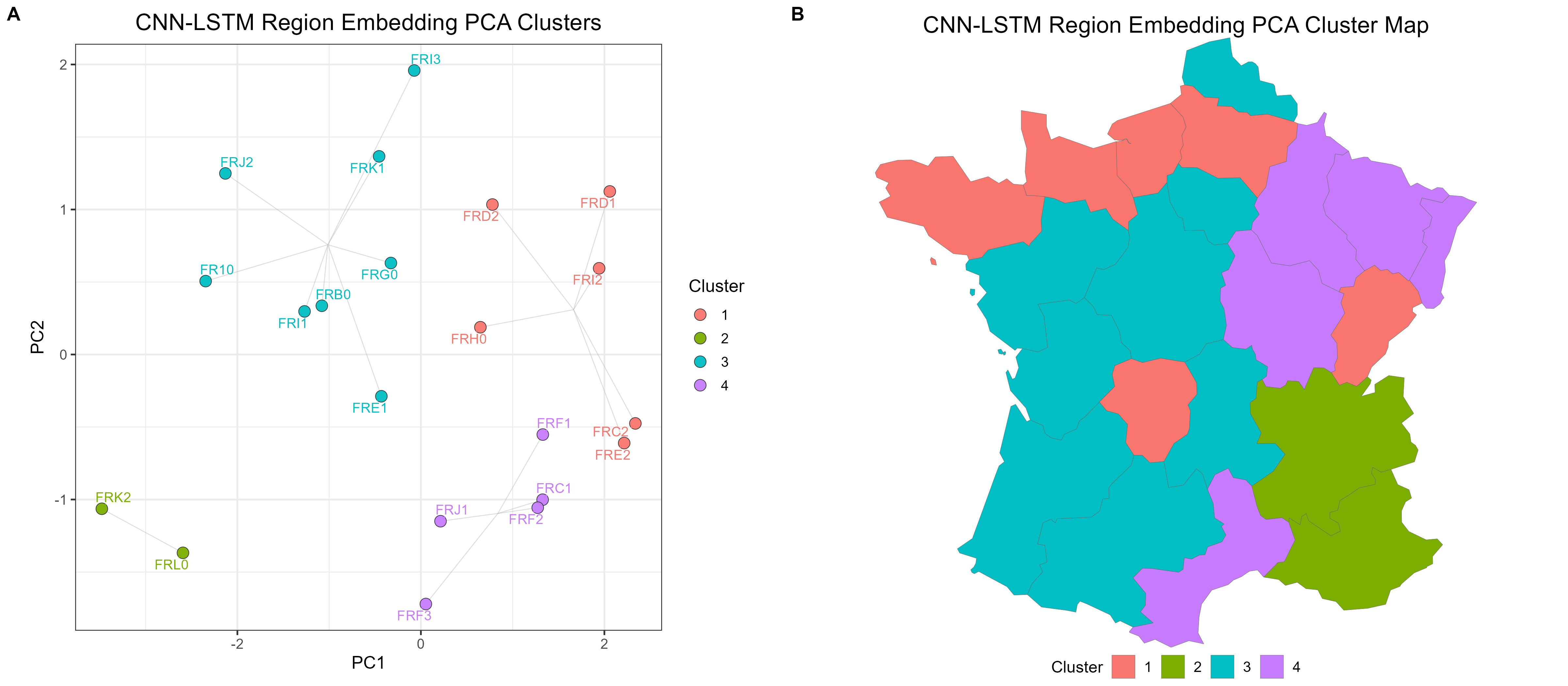}
    \caption{CNN--LSTM regional embedding clusters. \textbf{(A)}~PCA projection of the trained regional embedding vectors. Each point represents one NUTS 2 region; colours denote $k$-means cluster assignment ($k = 4$); line segments connect each region to its cluster centroid. \textbf{(B)}~Geographic map of cluster assignments across French NUTS 2 regions.}
    \label{fig:embed_cnn}
\end{figure}

\begin{figure}[!htb]
    \centering
    \includegraphics[width=1\linewidth]{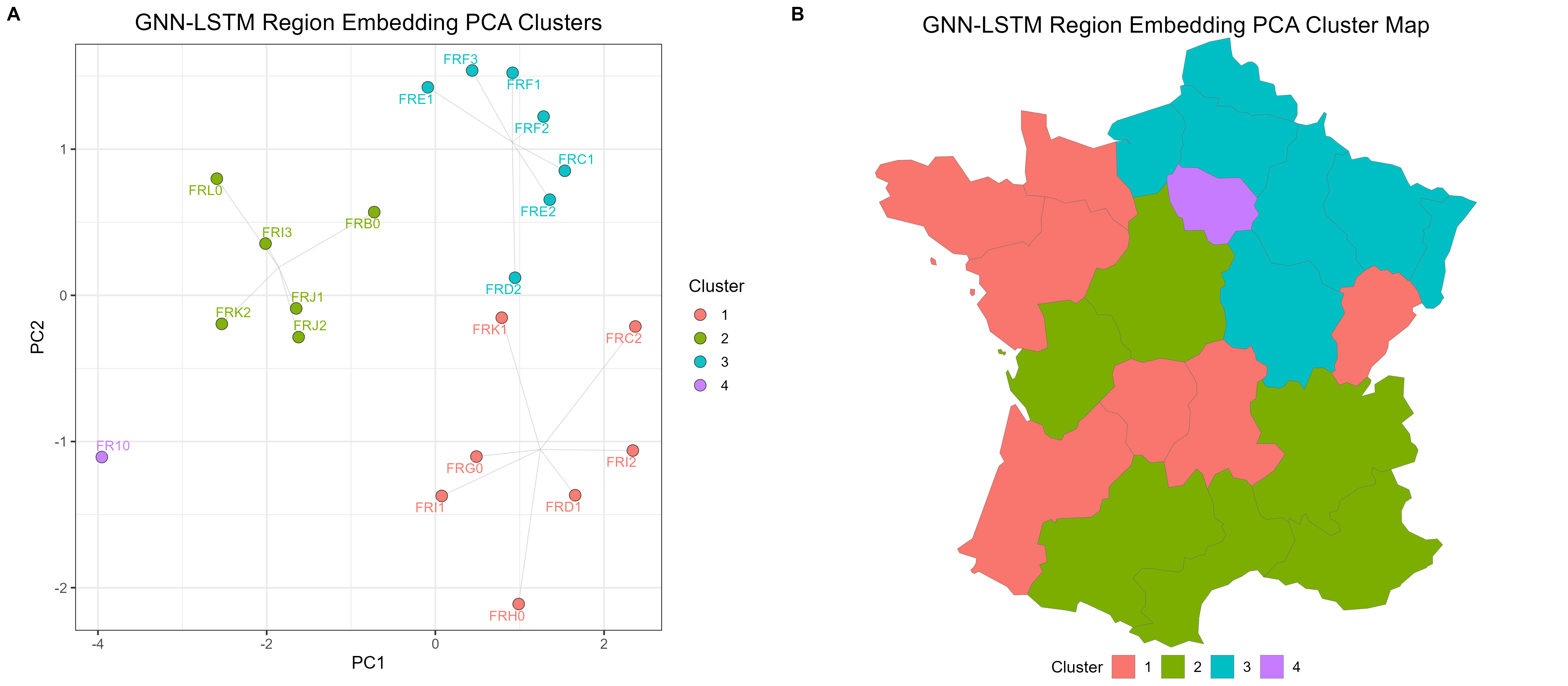}
    \caption{GNN--LSTM regional embedding clusters, using the same layout conventions as Figure~\ref{fig:embed_cnn}. \textbf{(A)}~PCA projection of the trained regional embedding vectors. \textbf{(B)}~Geographic map of cluster assignments across French NUTS 2 regions.}
    \label{fig:embed_gnn}
\end{figure}

For the CNN--LSTM embeddings in Figure~\ref{fig:embed_cnn}, cluster~1 encompasses much of the northwest. Cluster~2 is localized in the southeast. Cluster~3 spans the west and center of France, forming the largest contiguous block, while Cluster~4 is concentrated in the northeast. The GNN--LSTM embeddings can be seen in Figure~\ref{fig:embed_gnn}. Cluster~1 covers the western area; Cluster~2 is concentrated in the south; Cluster~3 spans the northeast; and Cluster~4 corresponds exclusively to Ile-de-France. The Ile-de-France forms an isolated singleton in the PCA projection, reflecting its distinct demographic profile relative to all other regions. 

\subsection{Permutation Feature Importance} 
\begin{figure}[tbhp]
    \centering
    
    \begin{minipage}[!htb]{0.49\textwidth}
        \centering
        \includegraphics[width=\linewidth]{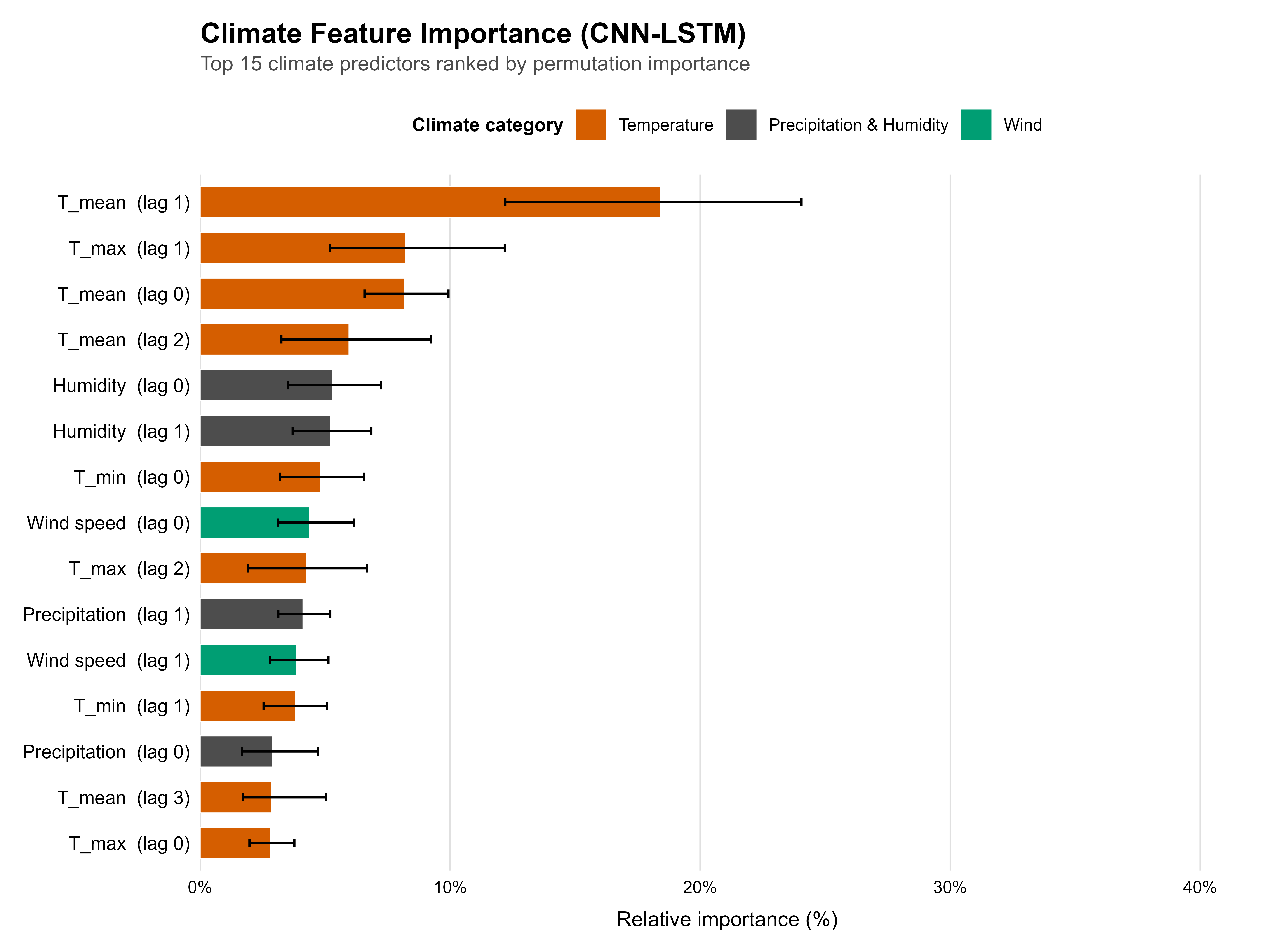}
    \end{minipage}
    \hfill
    \begin{minipage}[!htb]{0.49\textwidth}
        \centering
        \includegraphics[width=\linewidth]{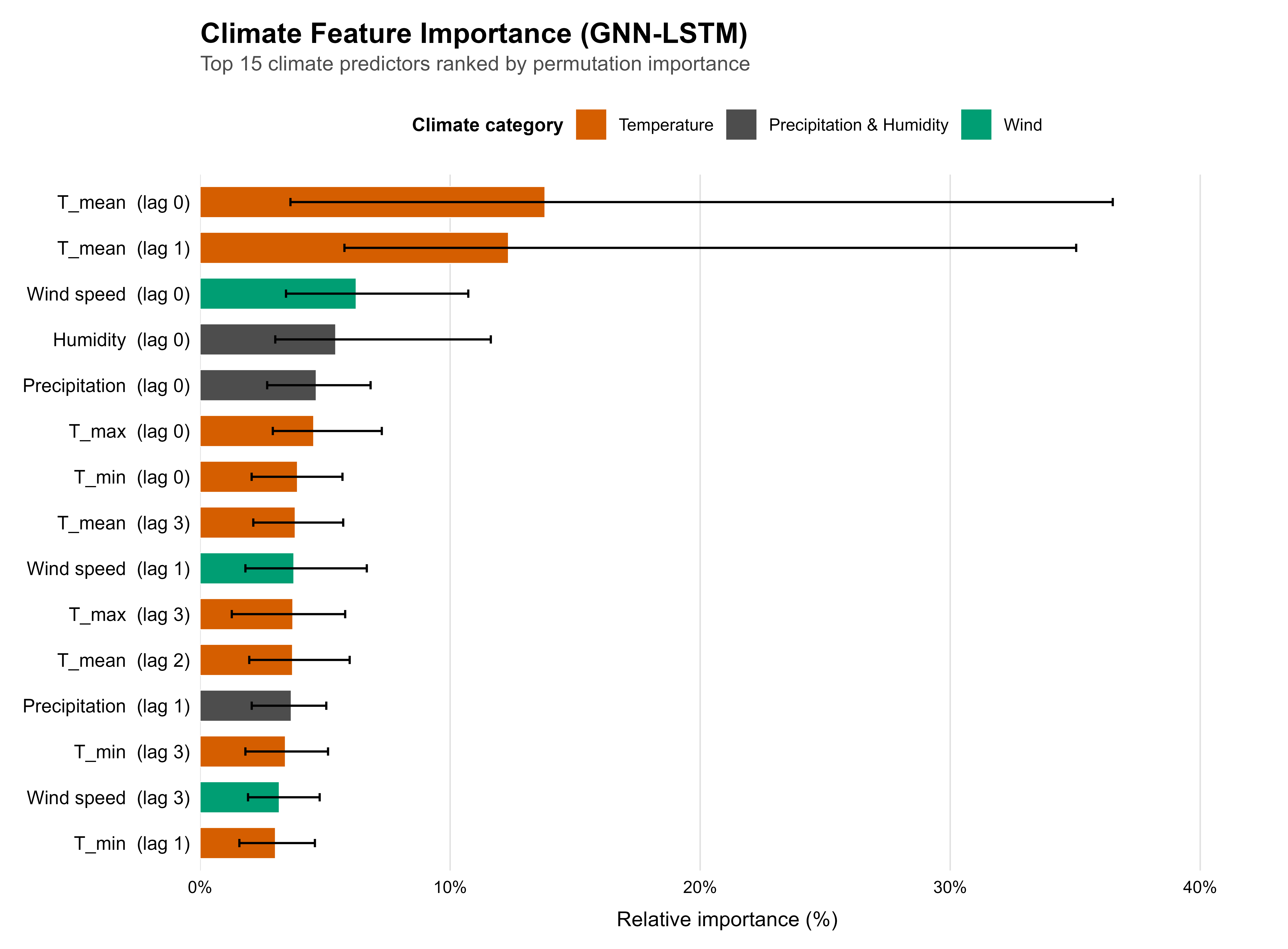}
    \end{minipage}
    \caption{Permutation feature importance of the top 15 climate predictors for the CNN--LSTM and GNN--LSTM models, estimated over 1,000 permutations. Relative importance is computed as the percentage increase in test MSE when each feature is shuffled, averaged across permutation samples. Error bars indicate one standard deviation across permutations.}
    \label{fig:climate_feature_importance}
\end{figure}

To assess which climate covariates drive excess mortality predictions, we compute permutation feature importance \citep{Breiman2001} for both the CNN--LSTM and GNN--LSTM. For each candidate feature, the values are randomly permuted 1{,}000 times across all observations in the test set, and the resulting increase in test MSE is recorded and expressed as a relative percentage. The reported importance values are averaged across permutation samples, while error bars reflect variability across draws. The results in Figure~\ref{fig:climate_feature_importance} reveal that temperature variables dominate in both models. For the CNN--LSTM, mean temperature ($T_{\mathrm{mean}}$) at Lag~1 is by far the most influential predictor, accounting for approximately 19\% of relative importance, nearly twice that of the next-ranked variable. Note that Lag $k$ refers to the climate observation $k$ weeks prior to the prediction. Next, mean temperature at Lag~0 and Lag~2 follow in second and third position, each contributing approximately 8--9\%. Collectively, the $T_{\mathrm{mean}}$ variables and their lags account for a substantial share of total importance, confirming that sustained temperature exposure across multiple preceding weeks is the primary driver of climate-induced excess mortality. Notably, the error bars for $T_{\mathrm{mean}}$ at Lag~1 are considerably wider than those of the remaining predictors, suggesting that the model's reliance on this feature varies across regions and age groups. The GNN--LSTM exhibits a similar importance profile. Mean temperature ($T_{\mathrm{mean}}$) at Lag~0 ranks first at approximately 15\%, followed closely by $T_{\mathrm{mean}}$ at Lag~1 at around 13\%, while wind speed at Lag~0 ranks third at roughly 7\%. As with the CNN--LSTM, the error bars for the most important temperature predictors are wider than those for lower-ranked features, reflecting heterogeneity in the sensitivity of individual regions and age groups to recent mean temperature. In both models, non-temperature variables such as precipitation, humidity, and wind speed individually account for approximately 3--7\% of total importance. This suggests that these variables capture secondary mechanisms, such as wind chill amplifying cold stress or elevated humidity increasing the effects of heat stress.

\subsection{Temperature--Mortality Response Curve} 
\begin{figure}[tbhp]
    \centering
    \begin{minipage}[!htb]{0.49\textwidth}
        \centering
        \includegraphics[width=\linewidth]{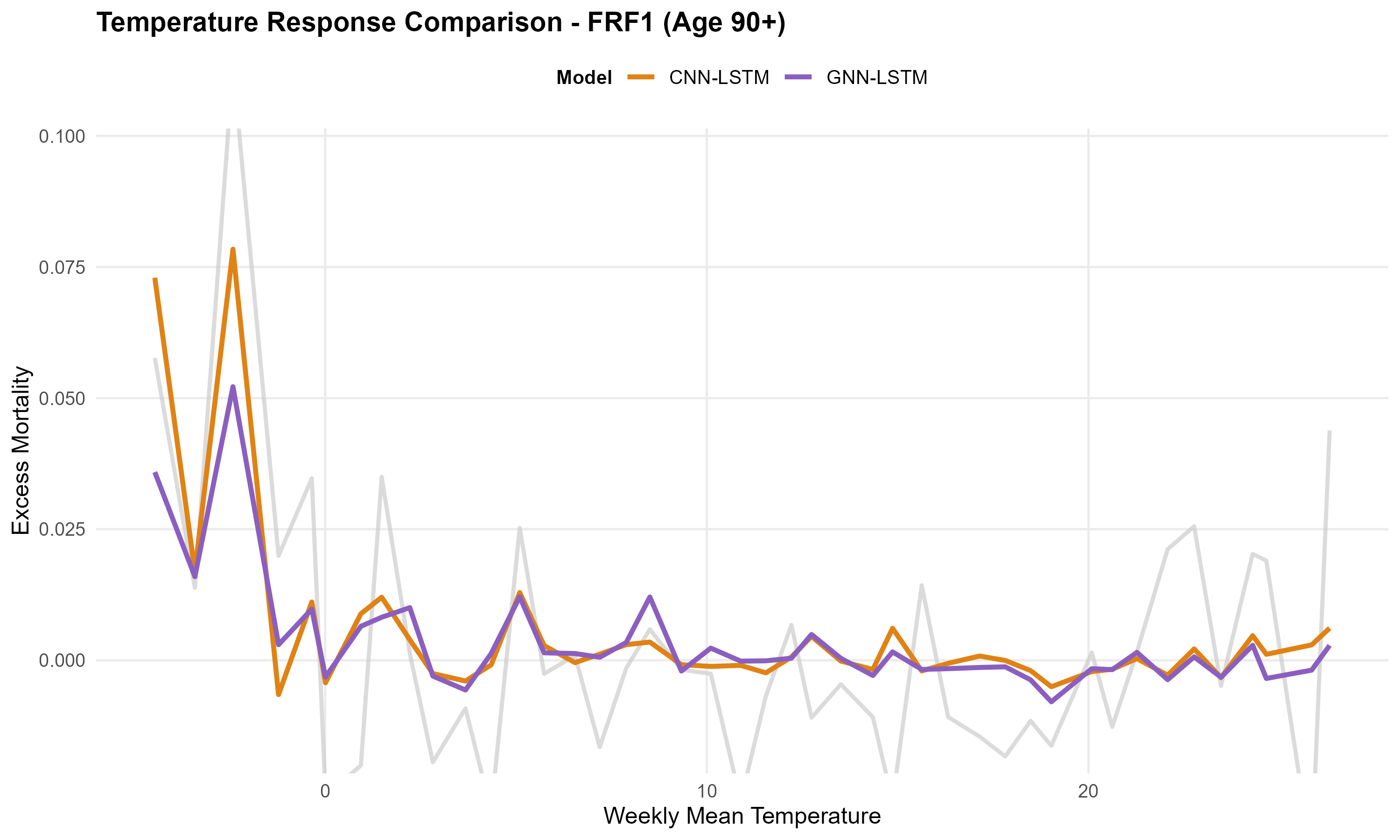}
    \end{minipage}
    \hfill
    \begin{minipage}[!htb]{0.49\textwidth}
        \centering
        \includegraphics[width=\linewidth]{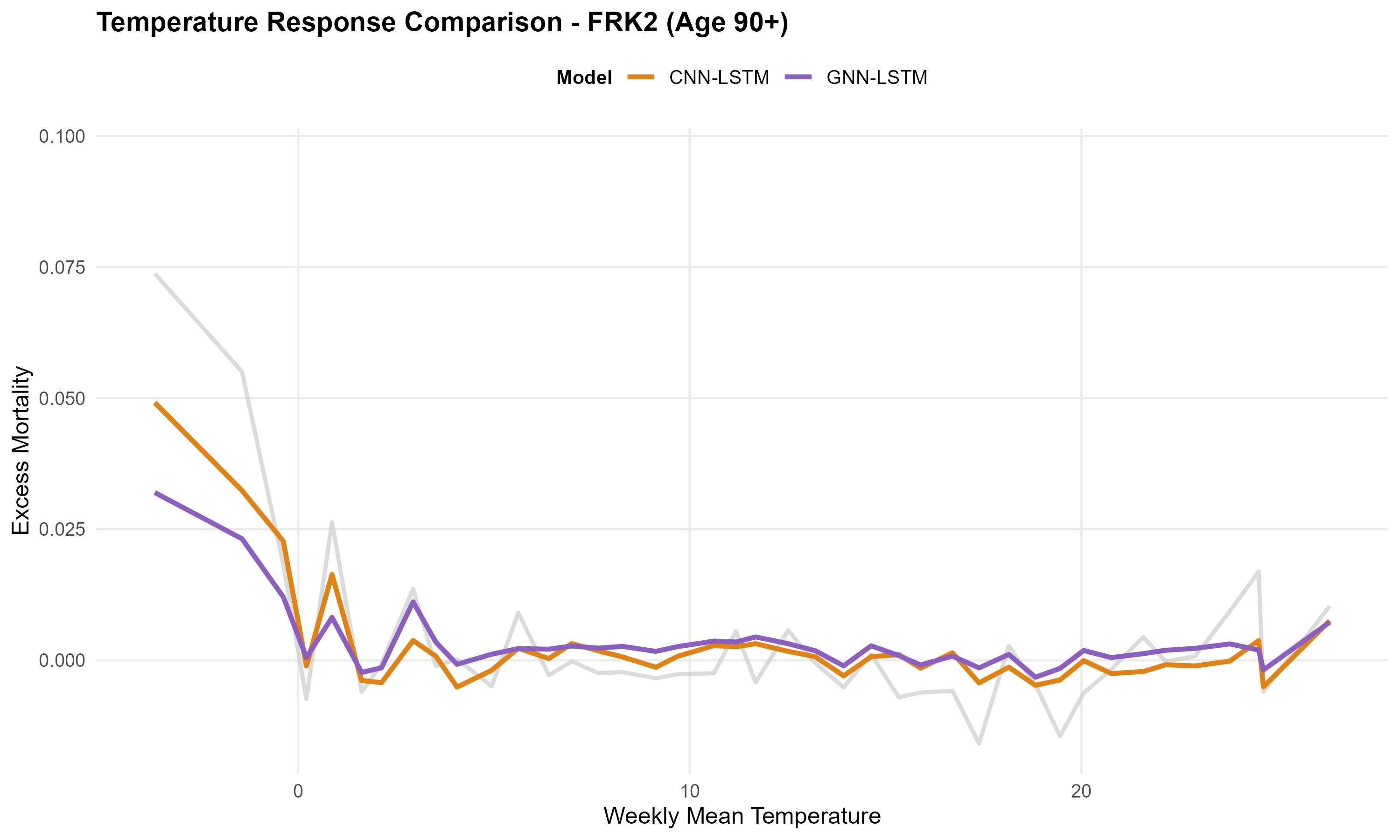}
    \end{minipage}
    \caption{Temperature response curves for the CNN--LSTM and GNN--LSTM in FRF1 and FRK2, for the age group 90+. Each curve plots predicted excess mortality against weekly mean temperature, with the grey line representing the median observed test-set excess mortality.}
    \label{fig:temperature_response}
\end{figure}

Figure~\ref{fig:temperature_response} illustrates the temperature--mortality response curves learned by the CNN--LSTM and GNN--LSTM for FRF1 and FRK2, focusing on the oldest age group. Both models capture the cold--mortality relationship, with predicted excess mortality increasing sharply as weekly mean temperature falls below approximately $5^\circ$C. Above approximately $10^\circ$C, predicted excess mortality remains close to zero in both regions, with only a slight increase at the upper end of the temperature range. This relatively flat heat response reflects the composition of the test set: the 2015--2019 period contains no major summer heat events, resulting in a negligible observed excess mortality signal at high temperatures. Consequently, the models learn only a weak heat-related response. 

\section{Uncertainty Quantification}
The total forecast uncertainty in the predicted log mortality rates arises from three sources: uncertainty in the estimated parameters of the baseline model, uncertainty in the projected national mortality trend and regional deviations, and uncertainty in the climate-driven residual predicted by the neural models. The first two sources concern the baseline; the third concerns the residual. Each source is quantified separately and then combined into a single predictive distribution. 

\subsection{Uncertainty in the Baseline}
The Lee--Carter baseline carries two sources of uncertainty that are 
propagated via simulation. The first is parameter uncertainty arising 
from the estimation of $\alpha_{a,r}$, $\beta_a$, $\kappa_{t,r}$, 
$\gamma_a$, and $\lambda_{w,r}$ from a finite sample of observed deaths. 
We propagate this via a parametric bootstrap: for each of $B = 10{,}000$ 
bootstrap replicates, we simulate synthetic death counts from a negative 
binomial distribution with mean $\hat{d}_{a,t,w,r} = 
\hat{\mu}^{\text{base}}_{a,t,w,r} \cdot E_{a,t,w,r}$ and dispersion 
$\hat{\phi}_{a,r}$, re-estimate the full Lee--Carter model on the 
simulated counts, and recompute the baseline log-mortality surface over 
the forecast horizon. This yields $B$ trajectories that reflect sampling 
uncertainty in the estimated parameters.

The second source is forecast uncertainty in the national trend $K_{t}$ and the regional deviations $u_{t,r}$. The national index follows a random walk with drift, and its forecast distribution is simulated by drawing independent Gaussian innovations with standard deviation equal to that of the differenced series. The regional deviations follow independent AR(1) processes, one per region, with coefficients and innovation variances estimated by maximum likelihood. For each of the $B$ simulation paths, future values of $K_{t}$ and $u_{t,r}$ are drawn recursively, combined as $\kappa_{t,r} = K_{t} + u_{t,r}$, and substituted into the Lee--Carter formula to produce an alternative mortality trajectory over the test period.

\subsection{Residual Uncertainty via Quantile Regression}
The third source of uncertainty concerns the climate-driven residual $R_{a,t,w,r}$. The CNN--LSTM and GNN--LSTM architectures are each extended to a quantile variant by modifying the output layer to produce simultaneous estimates $\hat{R}_{a,t,w,r}^{(q)}$ at quantile levels $q \in \{0.05, 0.50, 0.95\}$, while the underlying architecture remains identical to the point-estimate case. We follow \cite{Wang2019quantile} such that the models are trained by minimizing the average pinball loss,
\begin{equation}
    \mathcal{L}_q(\boldsymbol{\theta})
    =
    \frac{1}{|\mathcal{D}_{\mathrm{train}}|}
    \sum_{(a,t,w,r)\in\mathcal{D}_{\mathrm{train}}}
    \ell_q\left(R_{a,t,w,r}, \hat{R}_{a,t,w,r}^{(q)}\right),
    \label{eq:pinball_loss}
\end{equation}
where the pinball loss for a single observation is
\begin{equation}
    \ell_q\left(R_{a,t,w,r}, \hat{R}_{a,t,w,r}^{(q)}\right)
    =
    \begin{cases}
        q \bigl(R_{a,t,w,r} - \hat{R}_{a,t,w,r}^{(q)}\bigr),
        & \text{if } R_{a,t,w,r} \geq \hat{R}_{a,t,w,r}^{(q)}, \\[0.5em]
        (1-q)\bigl(\hat{R}_{a,t,w,r}^{(q)} - R_{a,t,w,r}\bigr),
        & \text{otherwise.}
    \end{cases}
    \label{eq:pinball}
\end{equation}
Minimizing Eq.~\eqref{eq:pinball_loss} yields an estimate of the conditional $q$-quantile of $R_{a,t,w,r}$. The predicted spread varies across age groups, time, week, and regions, widening during periods of climatic stress and contracting under stable conditions. 

The three uncertainty sources are combined into a single predictive distribution for each cell $(a,t,w, r)$. The three predicted residual quantiles $\hat{R}^{(0.05)}$, $\hat{R}^{(0.50)}$, $\hat{R}^{(0.95)}$ are used to fit a split-normal distribution \citep{salem2020prediction}, parameterized by median $\hat{R}^{(0.50)}$ and asymmetric standard deviations
\begin{equation*}
    \sigma_L = \frac{\hat{R}^{(0.50)} - \hat{R}^{(0.05)}}{\Phi^{-1}(0.95)},
    \qquad
    \sigma_R = \frac{\hat{R}^{(0.95)} - \hat{R}^{(0.50)}}{\Phi^{-1}(0.95)},
\end{equation*}
where $\Phi^{-1}$ denotes the standard normal quantile function. For each Monte Carlo draw $b$, a residual $R^{(b)}_{a,t,w,r}$ is sampled from this distribution and added to the bootstrapped baseline, giving
\begin{equation}
    \log\left(\hat{\mu}^{\text{final},(b)}_{a,t,w,r}\right)
    =
    \log\left(\hat{\mu}^{\text{base},(b)}_{a,t,w,r}\right)
    +
    R^{(b)}_{a,t,w,r}.
    \label{eq:combined_uq}
\end{equation}
where $\hat{\mu}^{\text{base},(b)}_{a,t,w,r}$ denotes the bootstrapped baseline for draw $b$, incorporating estimation and simulation uncertainty. Empirical quantiles of the resulting $B$-sample distribution of $\hat{\mu}^{\text{final},(b)}_{a,t,w,r}$ define the prediction interval $[\hat{\mu}^{\text{final},(0.05)}_{a,t,w,r}, \hat{\mu}^{\text{final},(0.95)}_{a,t,w,r}]$, targeting 90\% nominal coverage. Interval quality is evaluated using two complementary metrics as used in \cite{Schnurch2022interval}. The Prediction Interval Coverage Probability (PICP) measures the proportion of observed mortality rates falling within the band,
\begin{equation}
    \mathrm{PICP}
    =
    \frac{1}{|\mathcal{D}_{\mathrm{test}}|}
    \sum_{(a,t,w,r)\in\mathcal{D}_{\mathrm{test}}}
     \mathds{1}\left(
        \hat{\mu}^{\text{final},(0.05)}_{a,t,w,r}
        \leq
        \hat{\mu}^{\text{final}}_{a,t,w,r}
        \leq
        \hat{\mu}^{\text{final},(0.95)}_{a,t,w,r}
    \right),
    \label{eq:picp}
\end{equation}
where $\mathds{1}(\cdot)$ denotes the indicator function, equal to one if its argument holds and zero otherwise. The Mean Prediction Interval Width (MPIW) measures the average width of the interval on the mortality rate scale,
\begin{equation}
    \mathrm{MPIW}
    =
    \frac{1}{|\mathcal{D}_{\mathrm{test}}|}
    \sum_{(a,t,w,r)\in\mathcal{D}_{\mathrm{test}}}
    \Bigl(
        \hat{\mu}^{\text{final},(0.95)}_{a,t,w,r}
        -
        \hat{\mu}^{\text{final},(0.05)}_{a,t,w,r}
    \Bigr).
    \label{eq:mpiw}
\end{equation}
A well-calibrated model achieves a PICP close to the target while keeping MPIW as narrow as possible.

\subsection{Uncertainty Evaluation} 
Table~\ref{tab:picp_mpiw} reports the PICP and MPIW for each model over the test period. The LC-only baseline achieves a PICP of only 0.386, meaning it fails to contain the observed mortality rate in more than 60\% of weeks. Its narrow MPIW of 0.0084 means that the baseline treats mortality as nearly deterministic and cannot accommodate the volatility of climate-driven mortality. Both deep learning models correct this substantially. The CNN--LSTM reaches a PICP of 0.919, above the nominal 90\% target, while the GNN--LSTM achieves 0.905, also above the nominal target. The CNN--LSTM attains its higher coverage by producing wider intervals on average (MPIW~$= 0.0333$), whereas the GNN--LSTM maintains tighter bounds (MPIW~$=0.0285$) at a small cost to coverage. The GNN--LSTM thus achieves nearly equivalent coverage while keeping intervals 14\% narrower on average, resulting in tighter uncertainty bands around mortality projections.

\begin{table}[!tbh]
\centering
\renewcommand{\arraystretch}{1.2}
\begin{tabular}{lcc}
\toprule
\textbf{Model} & \textbf{PICP} & \textbf{MPIW} \\
\midrule
Baseline  & 0.3863 & \textbf{0.0084} \\
MortFCNet  & 0.8393 & 0.0239 \\
CNN--LSTM & \textbf{0.9192} & 0.0333 \\
GNN--LSTM & 0.9047 & 0.0285 \\
\bottomrule
\end{tabular}
\caption{Prediction interval coverage probability and mean prediction interval width for the Lee--Carter baseline and three neural network residual models, evaluated at the 90\% nominal coverage level over the training period 1990--2014 and testing period 2015--2019. A PICP close to 0.90 indicates well-calibrated uncertainty quantification.}
\label{tab:picp_mpiw}
\end{table}

\begin{figure}[!htb]
    \centering
    \includegraphics[width=\linewidth]{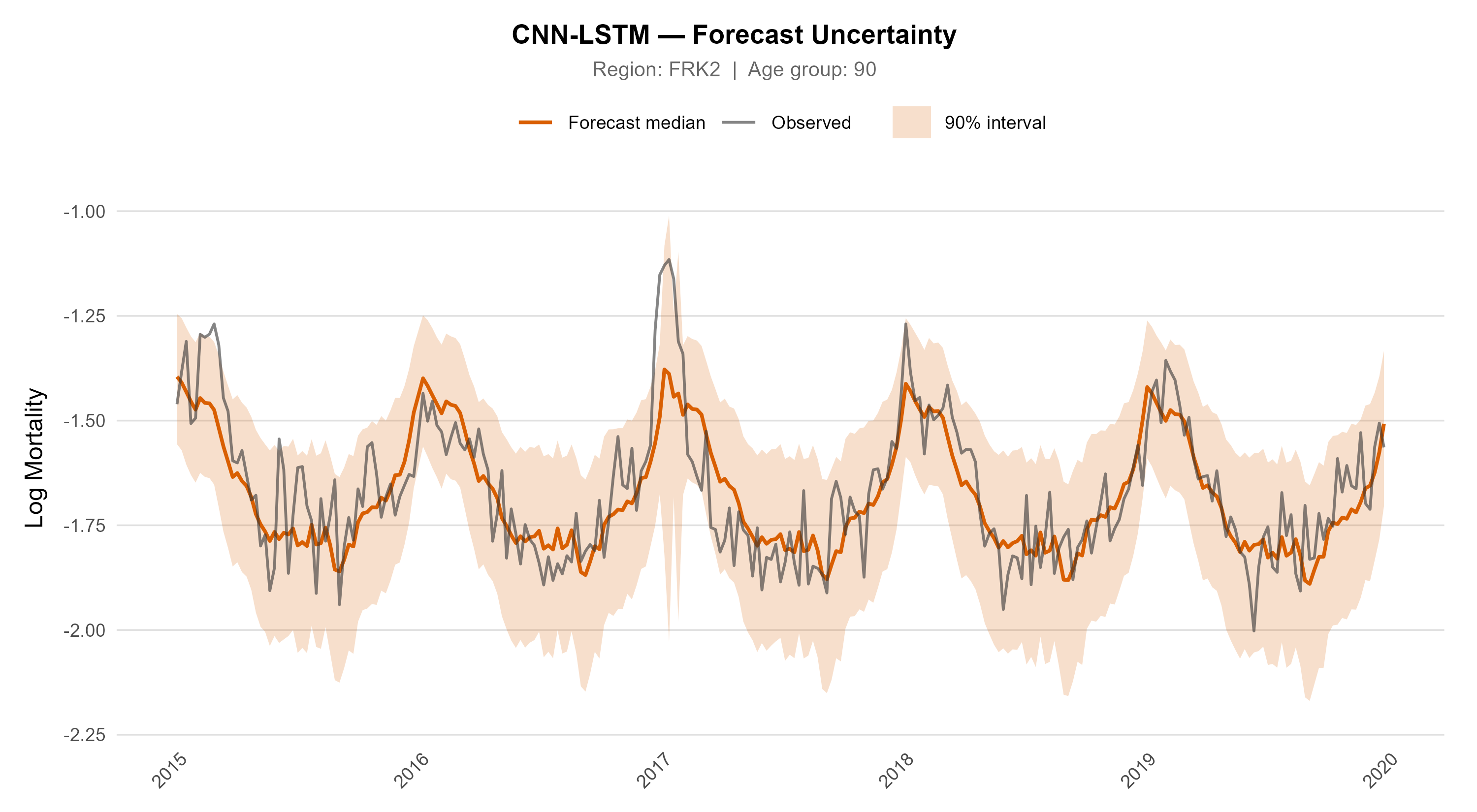}
    \caption{CNN--LSTM 90\% prediction interval for the log mortality rate in region FRK2, age group 90+, with training period 1990--2014 and test period 2015--2019. The shaded ribbon spans the $5\%$ and $95\%$ empirical quantiles of the combined predictive distribution, which aggregates three sources of uncertainty: Lee--Carter parameter uncertainty propagated via bootstrap, forecast uncertainty in the national trend and regional deviations, and residual uncertainty from the quantile LSTM.}
    \label{fig:uq_cnn}
\end{figure}

\begin{figure}[!htb]
    \centering
    \includegraphics[width=\linewidth]{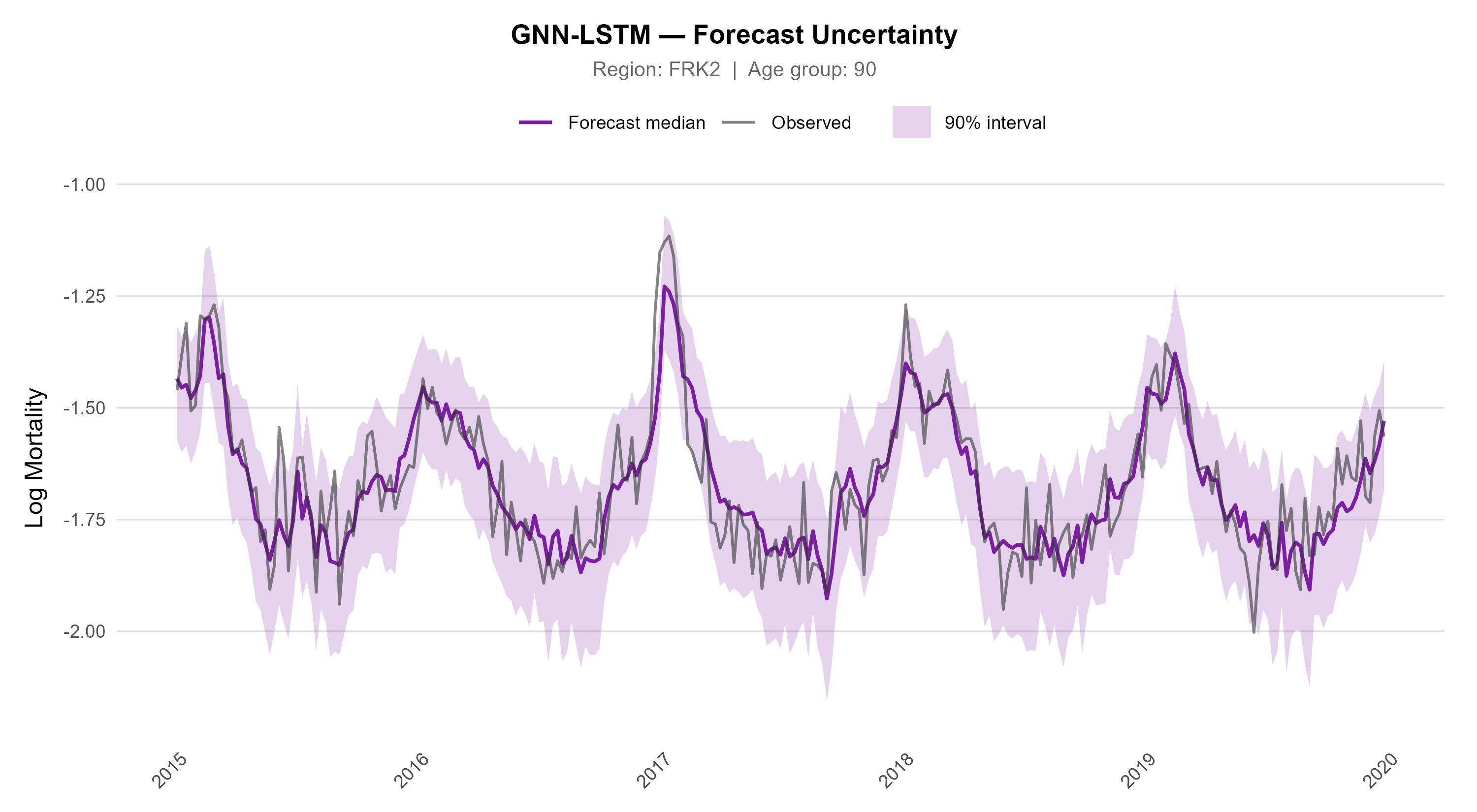}
    \caption{GNN--LSTM 90\% prediction interval for the log mortality rate in region FRK2, age group 90+, with training period 1990--2014 and test period 2015--2019. The shaded ribbon spans the $5\%$ and $95\%$ empirical quantiles of the combined predictive distribution, which aggregates three sources of uncertainty: Lee--Carter parameter uncertainty propagated via bootstrap, forecast uncertainty in the national trend and regional deviations, and residual uncertainty from the quantile LSTM.}
    \label{fig:uq_gnn}
\end{figure}

Figures~\ref{fig:uq_cnn} and~\ref{fig:uq_gnn} illustrate this behavior for region FRK2 and the 90+ age group. Both models capture the seasonal cycle, with the CNN--LSTM producing wider intervals throughout and the GNN--LSTM maintaining narrower bounds that nonetheless widen visibly around seasonal peaks, reflecting sensitivity to climatic stress. Both models contain the large majority of observed values across the test period, consistent with their empirical coverage rates exceeding the 90\% nominal target.

\section{Conclusion}
This paper presents a two-step climate-integrated mortality forecasting framework. We first established the weekly seasonal Lee--Carter model of \citet{robben2025penalizeddistributedlagnonlinear} as a baseline capturing both age-specific trends and intra-annual mortality dynamics, providing a foundation against which climate-driven deviations can be isolated across age groups and regions. Building on this baseline, we introduced two deep learning architectures trained on the residual mortality. The CNN--LSTM extends the MortFCNet framework of \citet{Zheng2025FineGrained} by incorporating explicit region embeddings and age representations, jointly modelling non-linear climate-mortality interactions across multiple climate variables, while the GNN--LSTM replaces standard convolutions with graph-based regional adjacency representations, enabling the model to learn spatial dependencies that reflect the heterogeneous climate regimes and demographic compositions of French NUTS 2 regions. Both architectures are further extended to a quantile LSTM framework that generates time-varying prediction intervals adapting to prevailing climate conditions. Both architectures outperform the Lee--Carter baseline and MortFCNet across nearly all regions, with gains increasing steadily with age and largest at ages 85 and above, where mortality is most sensitive to extreme climate events. 

The better performance of the CNN--LSTM and GNN--LSTM relative to MortFCNet indicates that the value of climate information depends on how it is incorporated into the forecasting model. MortFCNet applies a common mapping from climate variables to mortality across all age groups and regions. In contrast, the CNN--LSTM and GNN--LSTM explicitly model demographic and geographic heterogeneity through age and regional embeddings, with the GNN--LSTM further accounting for spatial dependencies between neighboring regions. This enables the models to learn population-specific responses to climatic conditions, resulting in improved predictive performance.

By decomposing observed mortality into a structural trend and a climate-driven residual, the framework provides a basis for separating how much of future mortality uncertainty stems from climate variability versus underlying demographic dynamics. For life insurers and pension funds, this decomposition provides insight into the extent to which mortality fluctuations are associated with climatic conditions rather than long-term demographic trends. In addition, the prediction intervals quantify how uncertainty in mortality forecasts changes over time and across regions, enabling a more informed assessment of climate-related mortality risk. As the frequency and severity of extreme climate events increase across Europe, frameworks that explicitly incorporate regional climate dynamics into mortality forecasting offer a more reliable foundation for the long-horizon projections that pension systems, insurers, and public health planners depend on.

\section*{Data and code availability statement}
All datasets employed in this study are publicly accessible. Mortality records were sourced from the INSEE ten-year files, which include information on all registered deaths in France since 1970, along with demographic population estimates by department, sex, and five-year age category (\href{https://www.insee.fr/fr/information/4769950}{https://www.insee.fr/fr/information/4769950} and \href{https://www.insee.fr/fr/statistiques/8331297}{https://www.insee.fr/fr/statistiques/8331297}. Daily climate observations were obtained from the E-OBS database distributed through the Copernicus Climate Data Store \href{https://cds.climate.copernicus.eu/datasets/insitu-gridded-observations-europe}{https://cds.climate.copernicus.eu/datasets/insitu-gridded-observations-europe}. The implementation code and related materials for this work are openly available at \href{https://github.com/kenrickso/Climate-Driven-Mortality-Forecasting-Using-Deep-Learning}{https://github.com/kenrickso/Climate-Driven-Mortality-Forecasting-Using-Deep-Learning}.

\bibliography{references}

\begin{thebibliography}{}

\bibitem[Ba et~al., 2016]{ba2016layernormalization}
Ba, J.~L., Kiros, J.~R., and Hinton, G.~E. (2016).
\newblock Layer normalization.
\newblock Working paper. \doi{10.48550/arXiv.1607.06450}.

\bibitem[Bai and Perron, 2003]{Bai2003Structural}
Bai, J. and Perron, P. (2003).
\newblock Computation and analysis of multiple structural change models.
\newblock {\em Journal of Applied Econometrics}, 18(1):1--22.
\newblock \doi{10.1002/jae.659}.

\bibitem[Barigou et~al., 2025]{barigou2025mortality}
Barigou, K., Patten, M., and Zhou, K.~Q. (2025).
\newblock Mortality modeling and forecasting with the actuaries climate index.
\newblock {\em arXiv preprint arXiv:2510.16266}.

\bibitem[Barnett et~al., 2010]{Barnett2010Temperature}
Barnett, A., Tong, S., and Clements, A. (2010).
\newblock What measure of temperature is the best predictor of mortality?
\newblock {\em Environmental Research}, 110(6):604–611.
\newblock \doi{10.1016/j.envres.2010.05.006}.

\bibitem[Braga et~al., 2002]{Braga2002Humidity}
Braga, A. L.~F., Zanobetti, A., and Schwartz, J. (2002).
\newblock The effect of weather on respiratory and cardiovascular deaths in 12 {U.S}. cities.
\newblock {\em Environmental Health Perspectives}, 110(9):859–863.
\newblock \doi{10.1289/ehp.02110859}.

\bibitem[Breiman, 2001]{Breiman2001}
Breiman, L. (2001).
\newblock Random forests.
\newblock {\em Machine Learning}, 45(1):5–32.
\newblock doi: 10.1023/a:1010933404324.

\bibitem[Bégin et~al., 2025]{Begin2025Seasonal}
Bégin, J.-F., Boudreault, M., and Landry, T. (2025).
\newblock Modelling seasonal mortality: An age–period–cohort approach.
\newblock {\em Insurance: Mathematics and Economics}, 125:103162.
\newblock \doi{10.1016/j.insmatheco.2025.103162}.

\bibitem[Bégin et~al., 2026]{Begin2026Seasonal}
Bégin, J.-F., Boudreault, M., and Landry, T. (2026).
\newblock Modelling the impacts of climate change on deaths caused by heat and cold waves with age-period-cohort models.
\newblock Working paper. \doi{10.13140/RG.2.2.20313.89442}.

\bibitem[Calabrese et~al., 2024]{Calabrese2024PhysicalInsurance}
Calabrese, R., Dombrowski, T., Mandel, A., Pace, R.~K., and Zanin, L. (2024).
\newblock Impacts of extreme weather events on mortgage risks and their evolution under climate change: A case study on {Florida}.
\newblock {\em European Journal of Operational Research}, 314(1):377–392.
\newblock \doi{10.1016/j.ejor.2023.11.022}.

\bibitem[Chen and Khaliq, 2023]{Chen2023LSTM}
Chen, Y. and Khaliq, A. Q.~M. (2023).
\newblock Mortality rates forecasting with data driven {LSTM, Bi-LSTM and GRU}: The {United States} case study.
\newblock {\em Actuarial Research Clearing House}, 2023(1).
\newblock \url{https://www.soa.org/globalassets/assets/files/static-pages/research/arch/2023/arch-2023-1-mortality-rates-forecasting.pdf}.

\bibitem[Chen et~al., 2019]{Chen2019DLNM}
Chen, Y.-H., Mukherjee, B., and Berrocal, V.~J. (2019).
\newblock Distributed lag interaction models with two pollutants.
\newblock {\em Journal of the Royal Statistical Society Series C: Applied Statistics}, 68(1):79--97.
\newblock \doi{10.1111/rssc.12297}.

\bibitem[{Copernicus Climate Change Service}, 2025]{eobs2025}
{Copernicus Climate Change Service} (2025).
\newblock {E-OBS} daily gridded meteorological data for {Europe} from 1950 to present derived from in-situ observations.
\newblock Dataset. DOI: 10.24381/cds.151d3ec6. Version v31.0e, accessed 2026-05-29.

\bibitem[Di~Febo, 2025]{DiFebo2025TransitionRisk}
Di~Febo, E. (2025).
\newblock Transition risk in climate change: A literature review.
\newblock {\em Risks}, 13(4):66.
\newblock \doi{10.3390/risks13040066}.

\bibitem[Enchev et~al., 2017]{Enchev2017Structural}
Enchev, V., Kleinow, T., and Cairns, A. J.~G. (2017).
\newblock Multi-population mortality models: fitting, forecasting and comparisons.
\newblock {\em Scandinavian Actuarial Journal}, 2017(4):319--342.
\newblock \doi{10.1080/03461238.2015.1133450}.

\bibitem[Gasparrini et~al., 2015]{Gasparrini2015dlnmImpact}
Gasparrini, A., Guo, Y., Hashizume, M., Lavigne, E., Zanobetti, A., Schwartz, J., Tobias, A., Tong, S., Rockl\"{o}v, J., Forsberg, B., Leone, M., De~Sario, M., Bell, M.~L., Guo, Y.-L.~L., Wu, C.-f., Kan, H., Yi, S.-M., de~Sousa Zanotti Stagliorio~Coelho, M., Saldiva, P. H.~N., Honda, Y., Kim, H., and Armstrong, B. (2015).
\newblock Mortality risk attributable to high and low ambient temperature: a multicountry observational study.
\newblock {\em The Lancet}, 386(9991):369–375.
\newblock \doi{10.1016/s0140-6736(14)62114-0}.

\bibitem[Giglio et~al., 2021]{Giglio2021Transition}
Giglio, S., Kelly, B., and Stroebel, J. (2021).
\newblock Climate finance.
\newblock {\em Annual Review of Financial Economics}, 13(Volume 13, 2021):15--36.
\newblock \doi{10.1146/annurev-financial-102620-103311}.

\bibitem[Goes et~al., 2025]{goes2025bayesian}
Goes, J., Barigou, K., and Leucht, A. (2025).
\newblock Bayesian mortality modelling with pandemics: a vanishing jump approach.
\newblock {\em Journal of the Royal Statistical Society Series C: Applied Statistics}, 74(4):1150--1182.

\bibitem[Guibert et~al., 2025]{Guibert2025dlnm}
Guibert, Q., Pincemin, G., and Planchet, F. (2025).
\newblock Impact of climate change on mortality: An extrapolation of temperature effects based on time series data in {France}.
\newblock {\em International Journal of Forecasting}.
\newblock \doi{10.1016/j.ijforecast.2025.07.004}.

\bibitem[Hainaut, 2018]{Hainaut2018Autoencoder}
Hainaut, D. (2018).
\newblock A neural-network analyzer for mortality forecast.
\newblock {\em ASTIN Bulletin}, 48(02):481–508.
\newblock \doi{10.1017/asb.2017.45}.

\bibitem[Hainaut, 2026]{Hainaut2026Socio}
Hainaut, D. (2026).
\newblock Explaining regional mortality differences with an economic-neural model: Evidence from {European NUTS-2} regions.
\newblock {\em LIDAM Discussion Paper}.
\newblock Working paper. \url{https://drive.google.com/file/d/1hJku2vojXti0AYkkkLrsQ2xnqXXjlGar/view}.

\bibitem[Jo et~al., 2022]{HyeongChan2022Quantile}
Jo, H., Kim, J., Huang, T.-C., and Ni, Y.-L. (2022).
\newblock {condLSTM-Q}: A novel deep learning model for predicting {COVID-19} mortality in fine geographical scale.
\newblock {\em Quantitative Biology}, 10(2):125--138.
\newblock \doi{10.15302/J-QB-021-0276}.

\bibitem[Kleinow, 2015]{Kleinow2015commonAgeEffect}
Kleinow, T. (2015).
\newblock A common age effect model for the mortality of multiple populations.
\newblock {\em Insurance: Mathematics and Economics}, 63:147--152.
\newblock \doi{10.1016/j.insmatheco.2015.03.023}.

\bibitem[Li and Tang, 2022]{Li2022}
Li, H. and Tang, Q. (2022).
\newblock Joint extremes in temperature and mortality: A bivariate {POT} approach.
\newblock {\em North American Actuarial Journal}, 26(1):43--63.
\newblock \doi{10.1080/10920277.2020.1823236}.

\bibitem[Li and Lee, 2005]{Li2005ACF}
Li, N. and Lee, R. (2005).
\newblock Coherent mortality forecasts for a group of populations: An extension of the {L}ee-{C}arter method.
\newblock {\em Demography}, 42(3):575–594.
\newblock \doi{10.1353/dem.2005.0021}.

\bibitem[Li et~al., 2026a]{Li2026MixedData}
Li, R., Zhou, R., and Pitt, D. (2026a).
\newblock Beyond annual data: Mortality forecasting with mixed frequency data.
\newblock {\em Insurance: Mathematics and Economics}, 126:103172.
\newblock \doi{10.1016/j.insmatheco.2025.103172}.

\bibitem[Li et~al., 2026b]{Li2026Dynamic}
Li, R., Zhou, R., and Pitt, D. (2026b).
\newblock Dynamic mortality forecasting via mixed-frequency state-space models.
\newblock Working paper. \doi{10.48550/arXiv.2601.05702}.

\bibitem[Lindholm and Palmborg, 2022]{Lindholm2022LSTM}
Lindholm, M. and Palmborg, L. (2022).
\newblock Efficient use of data for {LSTM} mortality forecasting.
\newblock {\em European Actuarial Journal}, 12(2):749–778.
\newblock \doi{10.1007/s13385-022-00307-3}.

\bibitem[Maas et~al., 2013]{Maas2013LeakyReLU}
Maas, A.~L., Hannun, A.~Y., and Ng, A.~Y. (2013).
\newblock Rectifier nonlinearities improve neural network acoustic models.
\newblock In {\em Proceedings of the 30th International Conference on Machine Learning (ICML)}, volume~28 of {\em JMLR Workshop and Conference Proceedings}, Atlanta, Georgia, USA.
\newblock \url{https://ai.stanford.edu/~amaas/papers/relu_hybrid_icml2013_final.pdf}.

\bibitem[Marino et~al., 2022]{Marino2022LSTM}
Marino, M., Levantesi, S., and Nigri, A. (2022).
\newblock A neural approach to improve the {Lee-Carter} mortality density forecasts.
\newblock {\em North American Actuarial Journal}, 27(1):148–165.
\newblock \doi{10.1080/10920277.2022.2050260}.

\bibitem[Miao et~al., 2026]{Miao2026GBM}
Miao, Z., Li, H., and Chen, Y. (2026).
\newblock Gradient boosted multi-population mortality modelling with high-frequency data.
\newblock Working paper. \doi{10.48550/arXiv.2507.09983}.

\bibitem[Min et~al., 2025]{Min2025dlnmLC}
Min, J., Li, H., Nagler, T., and Li, S. (2025).
\newblock Assessing climate-driven mortality risk: A stochastic approach with distributed lag non-linear models.
\newblock Working paper. \doi{10.48550/arXiv.2506.00561}.

\bibitem[Perla et~al., 2024]{Perla2024Multitask}
Perla, F., Richman, R., Scognamiglio, S., and W\"{u}thrich, M.~V. (2024).
\newblock Accurate and explainable mortality forecasting with the {LocalGLMnet}.
\newblock {\em Scandinavian Actuarial Journal}, 2024(7):739–761.
\newblock \doi{10.1080/03461238.2024.2307620}.

\bibitem[Perla et~al., 2021]{Perla2021DeepLC}
Perla, F., Richman, R., Scognamiglio, S., and Wüthrich, M.~V. (2021).
\newblock Time-series forecasting of mortality rates using deep learning.
\newblock {\em Scandinavian Actuarial Journal}, 2021(7):572--598.
\newblock \doi{10.1080/03461238.2020.1867232}.

\bibitem[Richman and W\"{u}thrich, 2019]{Richman2019NNLC}
Richman, R. and W\"{u}thrich, M.~V. (2019).
\newblock A neural network extension of the {L}ee-{C}arter model to multiple populations.
\newblock {\em Annals of Actuarial Science}, 15(2):346–366.
\newblock \doi{10.1017/s1748499519000071}.

\bibitem[Robben et~al., 2026a]{Robbens2025weeklyEstimation}
Robben, J., Antonio, K., and Kleinow, T. (2026a).
\newblock The short-term association between environmental variables and mortality: evidence from europe.
\newblock {\em Journal of the Royal Statistical Society Series A: Statistics in Society}, 189(2):1131--1153.
\newblock \doi{10.1093/jrsssa/qnaf052}.

\bibitem[Robben and Barigou, 2025]{robben2025penalizeddistributedlagnonlinear}
Robben, J. and Barigou, K. (2025).
\newblock A penalized distributed lag non-linear {Lee-Carter} framework for regional weekly mortality forecasting.
\newblock Working paper. \doi{10.48550/arXiv.2509.24087}.

\bibitem[Robben et~al., 2026b]{robben2025granular}
Robben, J., Barigou, K., and Kleinow, T. (2026b).
\newblock Granular mortality modeling with temperature and epidemic shocks: A three-state regime-switching approach.
\newblock {\em Insurance: Mathematics and Economics}, 128:103250.
\newblock \doi{10.1016/j.insmatheco.2026.103250}.

\bibitem[Salem et~al., 2020]{salem2020prediction}
Salem, T.~S., Langseth, H., and Ramampiaro, H. (2020).
\newblock Prediction intervals: Split normal mixture from quality-driven deep ensembles.
\newblock In {\em Conference on Uncertainty in Artificial Intelligence}, pages 1179--1187. PMLR.

\bibitem[Schnürch and Korn, 2022]{Schnurch2022interval}
Schnürch, S. and Korn, R. (2022).
\newblock Point and interval forecasts of death rates using neural networks.
\newblock {\em ASTIN Bulletin}, 52(1):333–360.
\newblock \doi{10.1017/asb.2021.34}.

\bibitem[Shala and Schumacher, 2024]{Shala2024PhysicalRisk}
Shala, I. and Schumacher, B. (2024).
\newblock The impact of natural disasters on banks’ impairment flow – evidence from {Germany}.
\newblock {\em Journal of Climate Finance}, 6:100031.
\newblock \doi{10.1016/j.jclimf.2024.100031}.

\bibitem[Shen et~al., 2024]{Shen2024GNN}
Shen, Y., Yang, X., Liu, H., and Li, Z. (2024).
\newblock Advancing mortality rate prediction in {European} population clusters: integrating deep learning and multiscale analysis.
\newblock {\em Scientific Reports}, 14(1).
\newblock \doi{10.1038/s41598-024-56390-x}.

\bibitem[So et~al., 2025]{So2025}
So, K.~R., Cruz, S.~C., Marcella, E.~A., Briones, J., and Garces, L. P.~D. (2025).
\newblock Uncertainty in pricing and risk measurement of survivor contracts.
\newblock {\em Risks}, 13(2).
\newblock \doi{10.3390/risks13020035}.

\bibitem[Srivastava et~al., 2014]{Srivastava2014Dropout}
Srivastava, N., Hinton, G., Krizhevsky, A., Sutskever, I., and Salakhutdinov, R. (2014).
\newblock Dropout: A simple way to prevent neural networks from overfitting.
\newblock {\em Journal of Machine Learning Research}, 15(56):1929--1958.
\newblock \url{http://jmlr.org/papers/v15/srivastava14a.html}.

\bibitem[Van~Berkum et~al., 2016]{vanBerkum2016Structural}
Van~Berkum, F., Antonio, K., and Vellekoop, M. (2016).
\newblock The impact of multiple structural changes on mortality predictions.
\newblock {\em Scandinavian Actuarial Journal}, 2016(7):581–603.
\newblock \doi{10.1080/03461238.2014.987807}.

\bibitem[Wang et~al., 2021]{Wang2021CNN}
Wang, C.-W., Zhang, J., and Zhu, W. (2021).
\newblock Neighbouring prediction for mortality.
\newblock {\em ASTIN Bulletin}, 51(3):689–718.
\newblock \doi{10.1017/asb.2021.13}.

\bibitem[Wang et~al., 2023]{Wang2023Transformer}
Wang, J., Wen, L., Xiao, L., and Wang, C. (2023).
\newblock Time-series forecasting of mortality rates using transformer.
\newblock {\em Scandinavian Actuarial Journal}, 2024(2):109–123.
\newblock \doi{10.1080/03461238.2023.2218859}.

\bibitem[Wang et~al., 2019]{Wang2019quantile}
Wang, Y., Gan, D., Sun, M., Zhang, N., Lu, Z., and Kang, C. (2019).
\newblock Probabilistic individual load forecasting using pinball loss guided {LSTM}.
\newblock {\em Applied Energy}, 235:10--20.
\newblock \doi{10.1016/j.apenergy.2018.10.078}.

\bibitem[Wuthrich and Merz, 2023]{Wuthrich2023Book}
Wuthrich, M.~V. and Merz, M. (2023).
\newblock {\em Statistical Foundations of Actuarial Learning and its Applications}.
\newblock Springer International Publishing.
\newblock \doi{10.1007/978-3-031-12409-9}.

\bibitem[Zhang et~al., 2022]{Zhang2022LSTMCNN}
Zhang, N., Chen, H., and LIU, J. (2022).
\newblock Mortality forecasting using {LSTM-CNN} model.
\newblock {\em SSRN Electronic Journal}.
\newblock Working paper. \doi{10.2139/ssrn.4261735}.

\bibitem[Zheng et~al., 2025]{Zheng2025FineGrained}
Zheng, H., Wang, H., Zhu, R., and Xue, J.-H. (2025).
\newblock Fine-grained mortality forecasting with deep learning.
\newblock {\em Annals of Actuarial Science}, page 1–27.
\newblock \doi{10.1017/s1748499525100171}.

\bibitem[Zheng et~al., 2026]{Zheng2026Review}
Zheng, H., Wang, H., Zhu, R., and Xue, J.-H. (2026).
\newblock A brief review of deep learning methods in mortality forecasting.
\newblock {\em Annals of Actuarial Science}, 20(1):150–165.
\newblock \doi{10.1017/S1748499525100110}.

\end{thebibliography}
\end{document}